\documentclass[prd,showpacs,nofootinbib,preprintnumbers]{revtex4}

\usepackage{amsmath} \usepackage{graphicx} \usepackage{amsfonts}
\usepackage{array} \usepackage{amsthm} \usepackage{bm}

\usepackage{latexsym}
\usepackage{epic} \usepackage{eepic}
\usepackage{graphicx,psfrag }
\newcommand{\HH}{{\cal H}}
\newcommand{\X}{{  X}}

\newcommand{\e}{{\rm e}}
\newcommand{\bb}{\bibitem}

\newcommand{\lb}{\label}

\newcommand{\bw}{\begin{widetext}}
\newcommand{\ew}{\end{widetext}}
\newcommand{\be}{\begin{equation}}
\newcommand{\ee}{\end{equation}}
\newcommand{\bea}{\begin{eqnarray}}
\newcommand{\eea}{\end{eqnarray}}
\newcommand{\nn}{\nonumber}

\newcommand{\V}{{\bar v}}
\newcommand{\U}{{\bar u}}
\newcommand\tom{\tilde\omega}
\begin{document}
\begin{flushright}DTP-MSU/07-19,   LAPTH-1203/07
\end{flushright}

\title{$G_2$ generating technique for minimal $D=5$ supergravity
and black rings}
\author{Adel Bouchareb} \email{bouchare@lapp.in2p3.fr}
\affiliation{Laboratoire de  Physique Th\'eorique LAPTH (CNRS),
B.P.110, F-74941 Annecy-le-Vieux cedex, France}

\author{Chiang-Mei Chen} \email{cmchen@phy.ncu.edu.tw}
\affiliation{Department of Physics, National Central University,
Chungli 320, Taiwan}

\author{G\'erard Cl\'ement} \email{gclement@lapp.in2p3.fr}
\affiliation{Laboratoire de  Physique Th\'eorique LAPTH (CNRS),
B.P.110, F-74941 Annecy-le-Vieux cedex, France}

\author{Dmitri V. Gal'tsov} \email{galtsov@phys.msu.ru}
\affiliation{Department of Theoretical Physics, Moscow State
University, 119899, Moscow, Russia}

\author{Nikolai G. Scherbluk} \email{shcherbluck@mail.ru}
\affiliation{Department of Theoretical Physics, Moscow State
University, 119899, Moscow, Russia}

\author{Thomas Wolf} \email{twolf@BrockU.ca}
\affiliation{Brock University, St.Catharines, Ontario, Canada L2S
3A1}

\date{\today}

\begin{abstract}
A solution generating technique is developed for $D=5$ minimal
supergravity with two commuting Killing vectors  based on the $G_2$
U-duality arising in the reduction of the theory to three
dimensions. The target space of the corresponding 3-dimensional
sigma-model  is the coset $G_{2(2)}/(SL(2,R)\times SL(2,R))$. Its
isometries constitute the set of solution generating symmetries.
These include  two electric and two magnetic Harrison
transformations  with the corresponding two pairs of gauge
transformations, three $SL(2,R) \; S$-duality transformations, and
the three gravitational scale, gauge and Ehlers transformations
(altogether 14). We construct a representation of the coset in terms
of $7\times 7$ matrices realizing the automorphisms of split
octonions. Generating a new solution amounts to transforming the
coset matrices by one-parametric subgroups of $G_{2(2)}$ and
subsequently solving the dualization equations. Using this formalism
we derive a new charged black ring solution with two independent
parameters of rotation.

\end{abstract}


\maketitle

\section{Introduction}

The discovery of rotating black rings \cite{er} (for a recent review see
\cite{er1} and references therein) has attracted  new  interest in
five-dimensional minimal supergravity \cite{cre,chani}. Within this
theory supersymmetric charged black ring solutions were found
\cite{gaun,elvasu}. The bosonic sector of five-dimensional minimal
supergravity is Einstein-Maxwell theory with a  Chern-Simons term,
the structure of the Lagrangian being similar to that of
eleven-dimensional supergravity \cite{mizo1,mizo2}. While in pure
Einstein-Maxwell theory in five and higher dimensions no charged black
hole solution, generalizing the uncharged Myers-Perry black holes
\cite{mype}, is known, the Chern-Simons
term endows five-dimensional Einstein-Maxwell theory with more
hidden symmetries, implying the existence of exact
charged rotating black hole solutions \cite{cve,gmt,kunz},
Meanwhile the most general black ring solution which might possess
mass, two angular momenta, electric charge and magnetic moment as
independent parameters is still not found. Here we propose a new
generating technique which can solve this problem. It is based on the
duality symmetries of the three-dimensional reduction of the theory.

The hidden symmetries arising upon dimensional reduction of
five-dimensional minimal supergravity to three dimensions were
studied by Mizoguchi and Ohta \cite{mizo1}, by Cremmer, Julia, Lu
and Pope \cite{cjlp} using the technique of \cite{cjlp1}, and were
more recently investigated both in the bosonic and fermionic sectors
by Possel \cite{pos1} (see also \cite{pos}). The corresponding
classical U-duality group  is the non-compact version of the lowest
rank exceptional group $G_2$ \cite{Gilmore}. In three dimensions one
obtains the gravity-coupled sigma-model with the homogeneous target
space $G_{2(2)}/SO(4)$ for the Lorentzian signature of the 3-space,
or $G_{2(2)}/(SL(2,R)\times SL(2,R))$ in the Euclidean case. Some
further aspects of these symmetries were discussed in \cite{mizo2},
their infinite-dimensional extension upon reduction to two and one
dimensions was also explored \cite{mizo3}.

Here we investigate the $G_{2(2)}/(SL(2,R)\times SL(2,R))$ sigma
model in the context of the solution generating technique  which has
proved to be extremely useful in various non-linear theories from
pure gravity, Einstein-Maxwell theory \cite{eh,ms,ern,nk,mg} and
dilatonic gravity \cite{gk,g,diak,cg,gl,gsh,ch} to more general
supergravity models \cite{ju,bgm,bm} and string theory \cite{sen2}.
Some partial use of hidden symmetries of this kind to generate new
rotating rings recently became a rapidly developing  industry. One
direction was to use the $SL(2,R)$ subgroup of the U-duality group
\cite{elva2,yaza}. Another line was related to the purely
gravitational sector (without the Maxwell field) which leads to
$SL(3, R)$ U-duality in three dimensions \cite{ms,beruf,har1}.
Further reduction to two dimensions gives rise to a
Belinsky-Zakharov type integrable model which was extensively used
to construct soliton solutions
\cite{har1,im,tn,ko,pome1,ak,mi,tmy,ef,posen,ekri,yaza1}. However,
the full $G_2$ symmetry was never used for generating purposes for
lack of a convenient representation of the coset
$G_{2(2)}/(SL(2,R)\times SL(2,R))$ in terms of the target space
variables. Although the 14-dimensional (adjoint) representation was
given explicitly in \cite{mizo1}, it is still too complicated for
practical generating applications. Here we construct a  suitable
representation in terms of $7\times 7$ matrices and give two
examples of its application: a sigma-model construction of the
electrically charged rotating black hole  and the generation of a
non-BPS doubly rotating charged black ring from the black ring with
two angular momenta of \cite{posen}.

Five-dimensional minimal supergravity contains a graviton, two $N=2$
symplectic-Majorana gravitini (equivalent to a single Dirac
gravitino), and one $U(1)$ gauge field. The bosonic part of the
Lagrangian is very similar to that of $D=11$ supergravity, being
endowed with a Chern-Simons term \cite{cre,chani}:
\begin{equation}
S_5 = \frac1{16 \pi G_5} \left[ \int d^5x \sqrt{- \hat g} \left(
\hat R - \frac14 \hat F^2 \right) - \frac1{3 \sqrt3} \int \hat F
\wedge \hat F \wedge \hat A \right],
\end{equation}
where $\hat F = d \hat A$.  This theory can be obtained  as a
suitably truncated Calabi-Yau compactification of $D=11$
supergravity~\cite{dimred}.

Our purpose is to construct a generating technique for classical
solutions with two commuting Killing symmetries. Dimensional
reduction leads to a three-dimensional sigma-model possessing a
$G_{2(2)}$ target space symmetry \cite{mizo1}. To explore it fully
we need a convenient representation of the action of symmetries on
the target space variables. We give here an alternative derivation
of the three-dimensional sigma-model which has the advantage of
being more explicit and easy to use for solution generation. The
reduction is performed in Sect. 2 in two steps, first to four, then
to three dimensions. In Sect. 3 we reveal the symmetries of the
three-dimensional sigma-model using a direct (computer assisted)
solution of the corresponding Killing equations\footnote{A purely
algebraic construction of the Killing vectors will be presented
elsewhere \cite{5to3}.}. The resulting symmetry transformations  are
identified in the usual terms of gauge, S-duality and
Harrison-Ehlers sectors. Then we reformulate in Sect. 4 the
problem in terms of a covariant (with respect to the
two-Killing plane) reduction
which is more suitable for constructing the matrix representation,
and give the coset matrix representative as a symmetrical $7\times7$
matrix. In Sect. 5 we identify the charging transformation, and
apply it to the construction of the doubly rotating charged black
ring.

\section{Dimensional reduction}

\subsection{ D=4}
Assuming that the five-dimensional metric and the Maxwell field
$\hat A$ do not depend on a space-like coordinate $z$, we arrive at
the four-dimensional Einstein theory with two Maxwell fields,
a dilaton and an axion. We parametrize the five-dimensional interval and
the Maxwell one-form as
\begin{eqnarray}
ds_5^2 &=& \mathrm{e}^{-2\phi } (dz + C_\mu dx^\mu)^2 +
\mathrm{e}^{\phi} ds_4^2,\label{ds5}\\
\hat A &=& A_\mu dx^\mu + \sqrt{3}\kappa dz,
\end{eqnarray}
($\mu = 1\ldots 4$).
 The corresponding four-dimensional action reads
\begin{equation}
S_4 = \frac1{16 \pi G_4} \int d^4x \sqrt{-g} \left[ R - \frac32
(\partial \phi)^2 - \frac32 \mathrm{e}^{2\phi} (\partial \kappa)^2 -
\frac14 \mathrm{e}^{-3\phi} G^2 - \frac14\mathrm{e}^{-\phi} \tilde
F^2 - \frac12 \kappa F F^* \right],
\end{equation}
where
\begin{equation}
G_4 = G_5/2\pi R_5, \quad G = d C, \quad F = d A, \quad \tilde F = F
+ \sqrt{3} C \wedge d\kappa,
\end{equation} and $F^*$ is the four-dimensional Hodge dual of $F$.

The dilaton $\phi$ and the axion $\kappa$ parametrize the coset
$SL(2,R)/U(1)$. To reveal the $SL(2,R)\; S$-duality symmetry in the
sector of vector fields ($A$ is inherited from 5D theory, $C$ is the
Kaluza-Klein vector) one has to reparametrize them using some
dualization \cite{mizo2}. We will reveal S-duality later on the
level of the further 3D reduction.

The field equations in terms of the four-dimensional variables read
\begin{eqnarray}
\nabla^2 \phi - \mathrm{e}^{2\phi} (\partial \kappa)^2 + \frac14
\mathrm{e}^{-3\phi} G^2 + \frac1{12} \mathrm{e}^{-\phi} \tilde F^2
&=& 0,
\\
\nabla_\mu \left( \mathrm{e}^{2\phi} \nabla^\mu \kappa \right) -
\frac1{3} \left[ \sqrt3\,  \nabla_\mu (\mathrm{e}^{-\phi} \tilde
F^{\mu\nu} C_\nu) + \frac12 F_{\mu\nu} F^{*\mu\nu} \right] &=& 0,
\\
\nabla_\mu \left( \mathrm{e}^{-\phi} \tilde F^{\mu\nu} + 2\kappa
F^{*\mu\nu} \right) &=& 0,
\\
\nabla_\mu \left( \mathrm{e}^{-3\phi} G^{\mu\nu} \right) +  \sqrt3\,
\mathrm{e}^{-\phi} \tilde F^{\mu\nu} \partial_\mu \kappa &=& 0,
\end{eqnarray}
and the Bianchi identities are
\begin{equation}\label{bian}
\nabla_\mu F^{*\mu\nu} = 0, \qquad \nabla_\mu G^{*\mu\nu} = 0.
\end{equation}
It is convenient to introduce the modified Maxwell tensors
\begin{eqnarray}
\mathcal{F}^{\mu\nu} &=& \mathrm{e}^{-\phi} \tilde F^{\mu\nu} + 2
\kappa F^{*\mu\nu},
\\
\mathcal{G}^{\mu\nu} &=& \mathrm{e}^{-3\phi} G^{\mu\nu} + \sqrt3\,
\left( \mathrm{e}^{-\phi} \kappa \tilde F^{\mu\nu} + \kappa^2
F^{*\mu\nu} \right),
\end{eqnarray}
in terms of which the Maxwell equations have the divergence form
\begin{equation} \label{maxw}
\nabla_\mu \mathcal{F}^{\mu\nu} = 0, \qquad \nabla_\mu
\mathcal{G}^{\mu\nu} = 0.
\end{equation}

\subsection{D=3}

Further reduction to $D=3$ is performed with respect to time,
assuming the standard parametrization of the stationary four-metric
\begin{equation}\label{ds4}
ds_4^2 = - f (dt - \omega_i dx^i)^2 + f^{-1} h_{ij} dx^i dx^j.
\end{equation}
The spatial part of the Bianchi identities (\ref{bian}) is solved
introducing the electric potentials $C_0 = \V_1, \, A_0 = \V_2$, so
that
\begin{equation}
G_{i0} = \partial_i \V_1, \qquad F_{i0} = \partial_i \V_2.
\end{equation}
Similarly,  the spatial components of the Maxwell equations
(\ref{maxw}) are solved by introducing magnetic potentials
$\U_1,\,\U_2$:
\begin{equation}
\mathcal{G}^{ij} = \frac{f}{\sqrt{h}} \, \epsilon^{ijk} \partial_k
\U_1, \qquad \mathcal{F}^{ij} = \frac{f}{\sqrt{h}} \, \epsilon^{ijk}
\partial_k \U_2.
\end{equation}
The time components of the corresponding equations then give  the
second order equations for these potentials. Straightforwardly we can
find (with the convention $\epsilon^{ijk} = - \epsilon^{0ijk}$)
\begin{eqnarray}
G^{ij} &=& \frac{f}{\sqrt{h}} \mathrm{e}^{3\phi} \, \epsilon^{ijk}
(w_1)_k, \qquad (w_1)_k := \partial_k \U_1 -   \sqrt3\, \,  \kappa
(\partial_k \U_2 - \kappa \partial_k \V_2),
\\
\tilde F^{ij} &=& \frac{f}{\sqrt{h}} \mathrm{e}^\phi \,
\epsilon^{ijk} (w_2)_k, \qquad (w_2)_k := \partial_k \U_2 - 2 \kappa
\partial_k \V_2,
\\
\tilde F_{i0} &=& (z_2)_i, \qquad\qquad\qquad\quad (z_2)_i :=
\partial_i \V_2 -  \sqrt3 \V_1 \partial_i \kappa.
\end{eqnarray}
The remaining components of the Maxwell tensors are obtained using
the following relations valid for any second rank four-dimensional
antisymmetric tensor $W_{\mu\nu}$ with the assumed form of the
metric (\ref{ds4}):
\begin{equation}
W^{i0} = W^{ij} \omega_j - h^{ij} W_{j0}, \qquad W_{ij} = f^{-2}
h_{ik} h_{jl} W^{kl} + 2 W_{0[i} \omega_{j]}.
\end{equation}
Using this we find:
\begin{eqnarray}
G^{i0} &=& \frac{f}{\sqrt{h}} \mathrm{e}^{3\phi} \, \epsilon^{ijk}
\omega_j (w_1)_k - \partial^i \V_1,
\\
G_{ij} &=& f^{-1} \sqrt{h} \mathrm{e}^{3\phi} \, \epsilon_{ijk}
(w_1)^k + 2 \omega_{[i} \partial_{j]} \V_1,
\\
\tilde F^{i0} &=& \frac{f}{\sqrt{h}} \mathrm{e}^{\phi} \,
\epsilon^{ijk} \omega_j (w_2)_k - (z_2)^i,
\\
\tilde F_{ij} &=& f^{-1} \sqrt{h} \mathrm{e}^{\phi} \,
\epsilon_{ijk} (w_2)^k + 2 \omega_{[i} (z_2)_{j]}, \end{eqnarray}
and for the squared quantities:
\begin{equation}
G^2 =  - 2 (\partial \V_1)^2 + 2 \mathrm{e}^{6\phi} (w_1)^2, \qquad
\tilde F^2  =  - 2 (z_2)^2 + 2 \mathrm{e}^{2\phi} (w_2)^2,
\end{equation}
where $(w_1)^2 = (w_1)_i (w_1)^i$.

Now we turn to the Einstein equations:
\begin{equation}\label{Ee}
R_{\mu\nu} - \frac32 \left( \partial_\mu \phi \partial_\nu \phi +
\mathrm{e}^{2\phi} \partial_\mu \kappa \partial_\nu \kappa \right) -
\frac12 \mathrm{e}^{-3\phi} \left( G_{\mu\alpha} G_\nu{}^\alpha -
\frac14 G^2 g_{\mu\nu} \right) - \frac12 \mathrm{e}^{-\phi} \left(
\tilde F_{\mu\alpha} \tilde F_\nu{}^\alpha - \frac14 \tilde F^2
g_{\mu\nu} \right) = 0.
\end{equation}
The Ricci tensor  decomposes as follows
\begin{eqnarray}\label{r00}
R_{00} &=& \frac12 \left( f \nabla^2 f - (\partial f)^2 + \tau^2
\right),
\\\label{r0i}
R^i{}_0 &=& - \frac{f}{2 \sqrt{h}} \, \epsilon^{ijk} \partial_j
\tau_k,
\\\label{rij}
R^{ij} &=& f^2 \mathcal{R}^{ij} - \frac12 \left[ \partial^i f
\partial^j f + \tau^i \tau^j \right] + h^{ij} R_{00},
\end{eqnarray}
where
\begin{equation}
\tau^i = -\frac{f^2}{\sqrt{h}} \, \epsilon^{ijk} \partial_j
\omega_k.
\end{equation}
The $0i$-part of (\ref{Ee}) can be solved introducing the twist
potential $\chi$ via
\begin{equation}
\tau_i = \partial_i \chi + \frac12 \left\{ \V_1 \partial_i \U_1 -
\U_1
\partial_i  \V_1  +    \V_2 \partial_i \U_2 - \U_2 \partial_i \V_2   +
\sqrt3\,  \left[ \kappa^2 \V_1 \partial_i \V_2 - \V_2 \partial_i
(\kappa^2 \V_1) \right] - \sqrt3\,  \left[ \kappa \V_1 \partial_i
\U_2 - \U_2
\partial_i (\kappa \V_1) \right]\right\}.
\end{equation}
Using this relation, we can rewrite the $00$-component of the Einstein
equations as
\begin{equation}
R_{00} = \frac14 f \left\{\mathrm{e}^{-3\phi} \left[ (\partial
\V_1)^2 + \mathrm{e}^{6\phi} (w_1)^2 \right] +  \mathrm{e}^{-\phi}
\left[ (z_2)^2 + \mathrm{e}^{2\phi} (w_2)^2 \right]\right\},
\end{equation}
and present the space-space part as
\begin{eqnarray}
R^{ij} &=& \frac32 f^2 h^{ia} h^{jb} \left( \partial_a \phi
\partial_b \phi + \mathrm{e}^{2\phi} \partial_a \kappa \partial_b
\kappa \right)
\nonumber\\
&-& \frac12 \mathrm{e}^{-3\phi} f \left[ \partial^i \V_1 \partial^j
\V_1 + \mathrm{e}^{6\phi} (w_1)^i (w_1)^j \right] + \frac14
\mathrm{e}^{-3\phi} f h^{ij} \left[ (\partial \V_1)^2 +
\mathrm{e}^{6\phi} (w_1)^2 \right]
\nonumber\\
&-& \frac12 \mathrm{e}^{-\phi} f \left[ (z_2)^i (z_2)^j +
\mathrm{e}^{2\phi} (w_2)^i (w_2)^j \right] +
\frac14\mathrm{e}^{-\phi} f h_{ij} \left[ (z_2)^2 +
\mathrm{e}^{2\phi} (w_2)^2 \right].
\end{eqnarray}
From the Eq. (\ref{r0i}) we then find that the Ricci tensor built on
the three-dimensional metric $h_{ij}$ will satisfy the following
three-dimensional Einstein equation
\begin{eqnarray}
\mathcal{R}_{ij} &=& \frac1{2f^2} \left( \partial_i f \partial_j f +
\tau_i \tau_j \right) + \frac32 \left( \partial_i \phi \partial_j
\phi + \mathrm{e}^{2\phi} \partial_i \kappa \partial_j \kappa
\right)
\nonumber\\
&-& \frac1{2f} \left[ \mathrm{e}^{-3\phi} \partial_i \V_1
\partial_j \V_1 + \mathrm{e}^{3\phi} (w_1)_i (w_1)_j
+  \mathrm{e}^{-\phi} (z_2)_i (z_2)_j +   \mathrm{e}^\phi (w_2)_i
(w_2)_j \right].
\end{eqnarray}
These equations derive from the action
\begin{equation}\label{3ds}
 S_3=\int \left(\mathcal{R}-{G}_{AB}\frac{\partial\Phi^A}{\partial x^i}
 \frac{\partial\Phi^B}{\partial x^j}h^{ij}\right)\sqrt{h}d^3x
\end{equation}
describing  the three-dimensional gravity coupled sigma model with
eight scalar fields
$\Phi^A=\{f,\chi,\phi,\kappa,\V_1,\V_2,\U_1,\U_2\}$ and the target
space metric $G_{AB}$ which will be given shortly. It can be checked
that the field equations for these quantities are equivalent to the
equations resulting from variation of this action over the $\Phi^A$.

In terms of the slightly rearranged potentials
\begin{equation}
v_1 := \V_1, \quad u_1 := \U_1 - \kappa^3 \V_1, \quad v_2 := \V_2 -
 \sqrt3\,   \kappa \V_1, \quad u_2 := \U_2 -  \sqrt3\,   \kappa^2 \V_1,
\end{equation}
 the equation for the twist potential $\chi$ will simplify to
\begin{equation}
\tau_i = \partial_i \chi + \frac12 ( v_1 \partial_i u_1 - u_1
\partial_i v_1   +  v_2 \partial_i u_2 - u_2
\partial_i v_2 ).
\end{equation}
The other variables will read
\begin{eqnarray}
(w_1)_i &:=& \partial_i u_1 + \kappa^3 \partial_i v_1 -  \sqrt3\,
\kappa (\partial_i u_2 - \kappa \partial_i v_2),
\nonumber\\
(w_2)_i &:=&  \left(\partial_i u_2 - 2 \kappa \partial_i v_2 -
 \sqrt3\,   \kappa^2 \partial_i v_1\right),
\\
(z_2)_i &:=&  \left(\partial_i v_2 +  \sqrt3\,   \kappa
\partial_i v_1\right).\nonumber
\end{eqnarray}

Finally, denoting $\xi=\ln f$ and shifting the twist potential
\begin{equation}
\chi\to\chi+\frac12(v_1u_1+v_2u_2),
\end{equation}
we can present the  metric of the  space of potentials
$\Phi^A=(\xi,\,\phi,\,\kappa,\,\chi,\,v_1,\,u_1,\,v_2,\,u_2)$ as
\begin{eqnarray}\label{metric}
dl^2 &=& {G}_{AB}d\Phi^{A}d\Phi^{B}= \frac12 \Biggl\{ d\xi^2 +
\mathrm{e}^{-2\xi} (d\chi + v_1 d u_1 + v_2 d u_2 ) ^2 + 3 \left(
d\phi^2 + \mathrm{e}^{2\phi} d\kappa^2 \right)
\nonumber\\
&-&  \mathrm{e}^{-\xi}\Biggl[ \mathrm{e}^{-3\phi} d v_1^2 +
\mathrm{e}^{3\phi} \left[ d u_1 + \kappa^3 d v_1 -   \sqrt3 \kappa
(d u_2 - \kappa d v_2) \right]^2
\nonumber\\
&& + \mathrm{e}^{-\phi} \left( d v_2 +  \sqrt3   \kappa d v_1
\right)^2 +  \mathrm{e}^\phi \left( d u_2 - 2 \kappa d v_2 -
 \sqrt3  \kappa^2 d v_1 \right)^2 \Biggr]\Biggr\}.
\end{eqnarray}
The target space is an eight-dimensional space with signature
$(++++----)$ similar to that of four-dimensional dilaton-axion
gravity with two Maxwell fields \cite{gsh}. In the latter case it
was identified as the $SU(2,2)/S(U(2)\times U(2))$ coset space
possessing a 15-dimensional isometry group and a 7-dimensional
isotropy subgroup. In our case the symmetry group is $G_{2(2)}$
\cite{mizo1,cjlp}. In what follows we present an independent
computer-assisted way to reveal the geometric structure of the space
(\ref{metric}).


\section{Isometries of the target space}
We used MAPLE  to calculate the Riemann tensor of the target space
and to check that all its covariant derivatives vanish, indicating
that it is a symmetric space. The set of Killing vectors was
identified using the CRACK code on REDUCE \cite{crack} to solve the
Killing equations for the metric (\ref{metric})
\begin{equation}
\X_{A;B} + \X_{B;A} = 0.
\end{equation}
The code gives 14 Killing vectors $\X_M$ (we use
$M,\,N\,\ldots=1,\ldots,14$ for numbering the Killing vectors and
$A,\,B\,\ldots=1,\ldots,8$  to denote the target space coordinates)
from which the first 9 are relatively simple:
\begin{eqnarray}
\X_1 &=& \partial_{\chi},\nonumber
\\
\X_2 &=& \partial_{u_1},\nonumber
\\
\X_3 &=& \partial_{u_2},\nonumber
\\
\X_4 &=& - u_1 \partial_\chi + \partial_{v_1},\nonumber
\\
\X_5 &=& - u_2 \partial_\chi + \partial_{v_2},
\\
\X_6 &=& \partial_\xi + \chi \partial_\chi + \partial_\phi - \kappa
\partial_\kappa + 2 v_1 \partial_{v_1} - u_1 \partial_{u_1} + v_2
\partial_{v_2},\nonumber
\\
\X_7 &=& 3 \partial_\xi + 3 \chi \partial_\chi + \partial_\phi -
\kappa
\partial_\kappa + 3 v_1 \partial_{v_1} + 2 v_2 \partial_{v_2} + u_2
\partial_{u_2},\nonumber
\\
\X_8 &=& v_2^2 \partial_\chi - \partial_\kappa - \sqrt3  u_2
\partial_{u_1} + \sqrt3  v_1 \partial_{v_2} - 2 v_2 \partial_{u_2},\nonumber\nonumber
\\
\X_9 &=&  {u_{2}}^{2}{\partial_\chi}+
{\kappa}^{2}{\partial_\kappa}-{e^{-2\,\phi}}{\partial_\kappa}-2\,
\kappa{\partial_\phi}-\sqrt {3}u_1 {\partial_{u_2}} +\sqrt {3}
v_2{\partial_{v_1}}-2\,u_2{\partial_{v_2}}.\nonumber
\end{eqnarray}
The remaining five are complicated in terms of the chosen
coordinates, we will present them in an alternative form later on.
The first eight Killing vectors span a Borel subalgebra, they are
used below to construct a matrix representation of the coset.

Already the number 14 of generators and the signature of the target
space indicate that one deals with the real non-compact form
$G_{2(2)}$ of the exceptional group $G_2$. The same code provides
the following list of commutation relations:
\begin{center}
\begin{tabular}{|c|c|c|c|c|c|c|c|c|c|c|}
\hline
  & $X_1$ & $X_2$ & $X_3$ & $X_4$ & $X_5$ & $X_6$ & $X_7$ & $X_8$ & $X_9$ & $X_{10}$ \\
\hline
  $X_1$ & 0 & 0 & 0 & 0 & 0 & $X_1$ & $3X_1$ & 0 & 0 & $3\X_5$ \\
  \hline
  $X_2$ &  & 0 & 0 & $-X_1$ & 0 & $-X_2$ & 0 & 0 & $-\sqrt3 X_3$ & $\sqrt3 X_8$ \\
  \hline
  $X_3$ &  &  & 0 & 0 & $-X_1$ & 0 & $X_3$ & $-\sqrt3 X_2$ & $-2X_5$ & $3X_6-2X_7$ \\
  \hline
  $X_4$ &  &  &  & 0 & 0 & $2X_4$ & $3X_4$ & $\sqrt3 X_5$ & 0 & 0 \\
  \hline
  $X_5$ &  &  &  &  & 0 & $X_5$ & $2X_5$ & $-2X_3$ & $\sqrt3 X_4$ & $2X_9$ \\
  \hline
  $X_{6}$ &  &  &  &  &  & 0 & 0 & $X_{8}$ & $-X_{9}$ & 0 \\
  \hline
  $X_{7}$ &  &  &  &  &  &  & 0 & $X_{8}$ & $-X_{9}$ & $X_{10}$ \\
  \hline
  $X_{8}$ &  &  &  &  &  &  &  & 0 & $3X_{6}-X_{7}$ & $-2X_{12}$ \\
  \hline
  $X_{9}$ &  &  &  &  &  &  &  &  & 0 & $-3X_{13}$ \\
  \hline
  $X_{10}$ &  &  &  &  &  &  &  &  &  & 0 \\
  \hline
\end{tabular}

\medskip

\begin{tabular}{|c|c|c|c|c|}
  \hline
    & $X_{11}$ & $X_{12}$ & $X_{13}$ & $X_{14}$ \\
  \hline
  $X_{1}$ & $X_{2}$ & $-3X_{3}$ & $\sqrt{3}X_{4}$ & $3X_{6}-3X_{7}$ \\
  \hline
  $X_{2}$ & 0 & 0 & $2\sqrt{3}X_{6}-\sqrt{3}X_{7}$ & $-3X_{11}$ \\
  \hline
  $X_{3}$ & 0 & $2X_{8}$ & $X_{9}$ & $X_{12}$ \\
  \hline
  $X_{4}$ & $X_{6}$ & $-\sqrt{3}X_{9}$ & 0 & $-\sqrt{3}X_{13}$ \\
  \hline
  $X_{5}$ & $\frac{\sqrt{3}}{3}X_{8}$ & $-X_{7}$ & 0 & $-X_{10}$ \\
  \hline
  $X_{6}$ & $2X_{11}$ & $X_{12}$ & $-X_{13}$ & $X_{14}$ \\
  \hline
  $X_{7}$ & $3X_{11}$ & $2X_{12}$ & 0 & $3X_{14}$ \\
  \hline
  $X_{8}$ & 0 & $-3\sqrt{3}X_{11}$ & $-X_{10}$ & 0 \\
  \hline
  $X_{9}$ & $-\frac{\sqrt{3}}{3}X_{12}$ & $-2X_{10}$ & 0 & 0 \\
  \hline
  $X_{10}$ & 0 & $3X_{14}$ & 0 & 0 \\
  \hline
  $X_{11}$ & 0 & 0 & $\frac{\sqrt{3}}{3}X_{14}$ & 0 \\
  \hline
  $X_{12}$ &  & 0 & 0 & 0 \\
  \hline
  $X_{13}$ &  &  & 0 & 0 \\
  \hline
  $X_{14}$ &  &  &  & 0 \\
  \hline
\end{tabular}
\end{center}
The  Killing metric constructed with the structure
constants gives the following set of non-zero scalar products
$(X_M,X_N) \equiv \eta_{MN}$:
 \begin{eqnarray}\label{kilme}
&(X_1,X_{14})=
\sqrt{3}(X_2,X_{13})=(X_3,X_{10})=-3(X_4,X_{11})=(X_5,X_{12})=
&\nonumber\\&= \frac32(X_6,X_{6}) =(X_6,X_{7})
=\frac12(X_7,X_{7})=(X_8,X_{9})=24. &
\end{eqnarray}
The Cartan subalgebra of $G_2$ is spanned by $X_6,\,X_7$, the basis
orthogonal with respect to the metric (\ref{kilme}) can be
chosen as
\begin{equation} H_1=X_7-\frac32 X_6,\quad H_2=\frac{\sqrt{3}}{2}
X_6.
\end{equation}
The remaining generators can be put in the Cartan-Weyl form
\begin{eqnarray}\label{cartan}
\left[H_i,E_{\bm{\alpha}}\right]&=&\alpha_i E_{\bm{\alpha}},\\
\left[E_{\bm{\alpha}},E_{-\bm{\alpha}}\right]&=&\alpha_i H_i,\\
\left[E_{\bm{\alpha}},E_{\bm{\beta}}\right]&=&N_{\bm{\alpha}\bm{\beta}}
E_{\bm{\alpha}+\bm{\beta}},
\end{eqnarray}
where $i=1,2$ and the roots $ \bm{\alpha}_{\pm a},\, a=1,\ldots,6$
are (with $H_1$ corresponding to the abscissa and $H_2$ to the
ordinate)
\begin{eqnarray}
 {\bm{\alpha}}_{\pm 1} &=&\pm(1,0), \nonumber\\
  {\bm{\alpha}}_{\pm 2} &=&\pm(-3/2,\sqrt{3}/2), \nonumber\\
  {\bm{\alpha}}_{\pm 3} &=&\pm(-1/2,\sqrt{3}/2), \nonumber\\
  {\bm{\alpha}}_{\pm 4} &=&\pm(1/2,\sqrt{3}/2), \\
  {\bm{\alpha}}_{\pm 5} &=&\pm(3/2,\sqrt{3}/2),\nonumber \\
  {\bm{\alpha}}_{\pm 6} &=&\pm(0,\sqrt{3}).\nonumber
\end{eqnarray}

\begin{figure}[!h]
\includegraphics[width=8cm,angle=0]{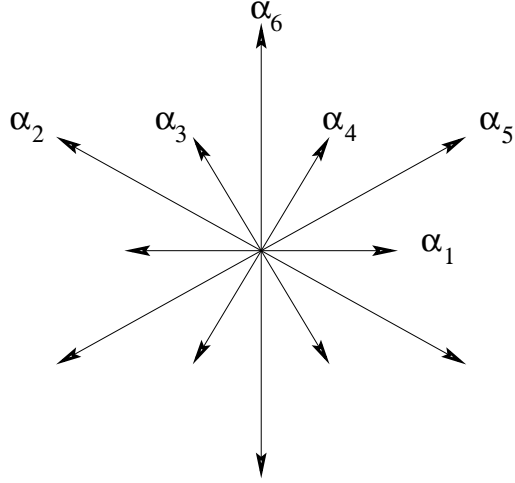}
\caption{The root diagram for $g_2$.}
\end{figure}

The simple roots are ${\bm{\alpha}}={\bm{\alpha}}_1$ and
${\bm{\beta}}={\bm{\alpha}}_2$, the remaining ones can be obtained
as follows:
\begin{equation}\label{root}
{\bm{\alpha}}_{\pm 3}=\pm({\bm{\alpha}}+{\bm{\beta}}),\
{\bm{\alpha}}_{\pm 4}=\pm(2{\bm{\alpha}}+{\bm{\beta}}),\
{\bm{\alpha}}_{\pm 5}=\pm(3{\bm{\alpha}}+{\bm{\beta}}),\
{\bm{\alpha}}_{\pm 6}=\pm(3{\bm{\alpha}}+2{\bm{\beta}}).
\end{equation}
The $E_{\bm{\alpha}}$ can be identified in terms of the $\X_M$ as
follows
\begin{eqnarray}
E_{1}&=&\sqrt{\frac16}X_{10},\quad\quad\quad\ \ E_{-1}=\sqrt{\frac32} X_{3},
\nonumber\\
E_{2}&=&\sqrt{\frac32} X_{2},\quad\quad\quad \;\;\ E_{-2}=\sqrt{\frac12}X_{13},
\nonumber\\
E_{3}&=&\sqrt{\frac12}\,X_{8},\quad\quad\quad \;\ E_{-3}=\sqrt{\frac12}\,X_{9},
\nonumber\\
E_{4}&=&\sqrt{\frac16}X_{12},\quad \quad\quad \;E_{-4}=\sqrt{\frac32}X_{5},
\\
E_{5}&=&\sqrt{\frac16}X_{14},\quad \quad\quad \;E_{-5}=\sqrt{\frac32}\,X_{1},
\nonumber\\
E_{6}&=&-\sqrt{\frac32}X_{11},\quad\;\;\; \;\ E_{-6}=\sqrt{\frac32}
X_{4}.\nonumber
\end{eqnarray}
The following structure constants are non-zero up to the standard
symmetries:
\begin{equation}
-N_{12}=N_{14}=N_{25}=N_{34}=\sqrt{\frac32},\quad N_{13}=\sqrt{2}.
\end{equation}

The symmetry of the root diagram under reflection implies the
existence of a target space coordinate transformation
$\Phi^A\to\tilde \Phi^{A }=(\tilde\xi ,\,\tilde\phi ,\,\tilde\kappa
,\,\tilde\chi ,\,\tilde v_1 ,\,\tilde u_1 ,\,\tilde v_2 ,\,\tilde
u_2 )$ such that
\begin{equation}
E_{-\bm{\alpha}}(\Phi)=E_{\bm{\alpha}}(\tilde \Phi ).
\end{equation}
In terms of the tilded variables the remaining 5 Killing vectors can
be written as (non-normalized)
\begin{eqnarray}\label{Ehars}
\X_{10} &=& \partial_{\tilde u_2},
\nonumber\\
\X_{11} &=& -\tilde u_1 \partial_{\tilde\chi}  + \partial_{\tilde
v_1},
\nonumber\\
\X_{12} &=& - \tilde u _2 \partial_{\tilde\chi}  +
\partial_{\tilde v_2},
\\
\X_{13} &=&\partial_{\tilde u_1},
\nonumber\\
\X_{14} &=& \partial_{\tilde\chi }.\nonumber
\end{eqnarray}

The target space isometries generated by the above Killing vectors can
be classified  as gauge, scale, S-duality and Harrison-Ehlers
transformations. The first, generated by $X_1$, is the gravitational
gauge transformation consisting in the shift of the twist potential:
\begin{equation}
  \chi \to \chi+\lambda_1;\ (\phi,\ \xi,\ v_1,\ v_2,\ u_1,\ u_2,\ \kappa\
  \hbox{invariant}),
\end{equation}
where $\lambda_1$ is the parameter. The next two, generated by
$X_2,\, X_3$, are magnetic gauge transformations shifting the
magnetic potentials $u_1$ (corresponding to the Kaluza-Klein vector
field) and $u_2$  (corresponding to the 5D Maxwell field):
\begin{eqnarray}
&u_1\to u_1+\lambda_2;\ (\phi,\ \xi,\ \chi,\ v_1,\ v_2,\ u_2,\
\kappa\ \hbox{invariant}),&\nonumber\\ &u_2\to u_2+\lambda_3;\
(\phi,\ \xi,\ \chi,\ v_1,\ v_2,\ u_1,\ \kappa\ \hbox{invariant}).&
\end{eqnarray}
The Killing vectors $X_4,\, X_5$ generate two electric gauge
transformations:
\begin{eqnarray}
&v_1\to v_1+\lambda_4,\ \chi\to \chi-u_1\lambda_4;\ (\phi,\ \xi,\
v_2,\ u_1,\ u_2,\ \kappa\ \hbox{invariant}),&\nonumber \\
&v_2\to v_2+\lambda_5,\ \chi\to \chi-u_2\lambda_5;\ (\phi,\ \xi,\
v_1,\ u_1,\ u_2,\ \kappa\ \hbox{invariant}).&
\end{eqnarray}
From  the sector $X_6,\,X_7$, one can separate the scale
transformation generated by $X_6-X_7$
\begin{eqnarray}
    \kappa &\to&\kappa,\quad u_1\to e^{\lambda_6}u_1,\quad v_1\to
    e^{\lambda_6} v_1,\quad \chi\to e^{2\lambda_6}\chi,\ \nonumber\\
    \phi &\to&\phi,\quad u_2\to e^{\lambda_6}u_2,\quad
    v_2  \to  e^{\lambda_6}v_2,\quad e^{\xi }\to e^{2\lambda_6}e^{\xi}.
\end{eqnarray}
The second independent element of the Cartan subalgebra
\begin{equation}\label{H}
   H=\frac12\left( \sqrt{3} H_2-H_1\right)= \frac12\bigg(3X_6-X_7\bigg),
\end{equation}
together with
\begin{equation}\label{L}
 L_-=X_9,\quad L_+=4\sqrt{3}\,X_8,
\end{equation}
form the  $SL(2,R)\, S$-duality  algebra
\begin{equation}\label{SL2}
\left[H, L_{\pm}\right]=\mp L_{\pm}, \quad  \left[L_-,
L_{+}\right]=H.
\end{equation}
The corresponding finite transformations consist of the  scaling
\begin{eqnarray}
    \chi &\to&\chi,\quad \kappa \to e^{-2\lambda_7}\kappa,\quad
    u_1 \to e^{-3\lambda_7}u_1\quad u_2\to e^{-\lambda_7}u_2\quad \nonumber\\
    \xi &\to&\xi,\quad e^{\phi }\to e^{2\lambda_7}e^{\phi}\quad
    v_1\to e^{3\lambda_7}v_1,\quad v_2\to e^{\lambda_7}v_2,
\end{eqnarray}
the axidilaton shift
\begin{eqnarray}
\kappa &\to&\kappa-\lambda_8, \nonumber
\\
\phi &\to&\phi, \nonumber
\\
u_1 &\to&-{\lambda_8}\,\sqrt{3} u_2+{{\lambda_8}}^{3}{
v_1}+{{\lambda_8}}^{2}\sqrt {3}{ v_2}+{ u_1},\nonumber
\\
u_2 &\to&{ u_2}-\sqrt {3}{{\lambda_8}}^{2}{ v_1}-2\,{\lambda_8}\,{
v_2},\nonumber
\\
v_1 &\to&v_1,
\\
v_2 &\to&{\lambda_8}\,\sqrt {3}{ v_1}+{ v_2},\nonumber
\\
\xi &\to&\xi,\nonumber
\\
\chi &\to&{{ v_1}}^{2}{{\lambda_8}}^{3}+\sqrt {3}{ v_2}\,{
v_1}\,{{\lambda_8}}^{2}+{{ v_2}}^{2}{\lambda_8}+\chi,\nonumber
\end{eqnarray}
and the shift of the inverted axidilaton
\begin{eqnarray}
\kappa &\to&{\frac {\kappa\,(1 -{\lambda_9}\,\kappa) -{\lambda_9
}\,{e^{-2\,\phi}}}{(1-{\lambda_9}\,\kappa)^{2}+{{
\lambda_9}}^{2}{e^{-2\,\phi}}}},\nonumber
\\
e^{\phi}&\to&\lambda_9^2\,e^{-\phi}+e^{\phi}(\lambda_9\,\kappa-1)^2,\nonumber
\\
u_1 &\to&u_1,
\\
u_2 &\to&-{\lambda_9}\,{ u_1}\,\sqrt {3}+{ u_2},\nonumber
\\
v_1 &\to&{\lambda_9}\,\sqrt {3}{ v_2}+{ v_1}-{{\lambda_9}}^{2}{u_2}
\,\sqrt {3}+{{\lambda_9}}^{3}{ u_1},\nonumber
\\
v_2 &\to&-2\,{ u_2}\,{\lambda_9}+{ v_2}+{{\lambda_9}}^{2}{ u_1}\,
\sqrt {3},\nonumber
\\
\xi &\to&\xi,\nonumber
\\
\chi &\to&{{\lambda_9}}^{3}{{ u_1}}^{2}-{{\lambda_9}}^{2}{
u_2}\,\sqrt { 3}{ u_1}+{\lambda_9}\,{{ u_2}}^{2}+\chi.\nonumber
\end{eqnarray}

The remaining generators $X_{10}, X_{11}, X_{12}, X_{13}, X_{14},$
form the Ehlers-Harrison sector. In terms of the tilded quantities
(\ref{Ehars}) the corresponding finite transformations are found
straightforwardly. They consist of two magnetic Harrison transformations
($X_{13},\,X_{10}$):
\begin{eqnarray}
&\tilde u_1 \to \tilde u_1+\tilde \lambda_{13};\ (\tilde \phi,\
\tilde \xi,\
\tilde \chi,\ \tilde v_1,\ \tilde v_2,\ \tilde u_2,\ \tilde \kappa\ \hbox{invariant}),&\nonumber\\
&\tilde u_2 \to \tilde u_2+\lambda_{10};\ (\tilde \phi,\ \tilde
\xi,\ \tilde \chi,\ \tilde v_1,\ \tilde v_2,\ \tilde u_1,\ \tilde
\kappa\ \hbox{invariant}),&
\end{eqnarray}
two electric Harrison transformations ($X_{11},\,X_{12}$):
\begin{eqnarray}
&\tilde v_1 \to \tilde v_1+\lambda_{11},\ \tilde \chi\to\tilde
\chi-u_1\lambda_{11};\ (\tilde \phi,\ \tilde \xi,\
\tilde v_2,\ \tilde u_1,\ \tilde u_2,\ \tilde \kappa\ \hbox{invariant}),&\nonumber\\
&\tilde v_2\to \tilde v_2+\lambda_{12},\ \tilde \chi\to\tilde
\chi-u_2\lambda_{12};\ (\tilde \phi,\ \tilde \xi,\ \tilde v_1,\
\tilde u_1,\ \tilde u_2,\ \tilde \kappa\ \hbox{invariant}),&
\end{eqnarray}
and an Ehlers transformation ($X_{14}$):
\begin{equation}
\tilde \chi \to \tilde \chi + \lambda_{14};\ (\tilde \phi,\ \tilde
\xi,\ \tilde v_1,\ \tilde v_2,\ \tilde u_1,\ \tilde u_2,\ \tilde
\kappa\ \hbox{invariant}).
\end{equation}

The above classification of the target space isometry
transformations is standard in dealing with four-dimensional
theories, where they have a particularly simple sense when applied
to stationary axisymmetric asymptotically flat configurations:
Harrison transformations are interpreted as charging
transformations, while the Ehlers transformation is interpreted as
generating a NUT parameter. In the five-dimensional setting this is
not so. Since one of the two four-dimensional vector fields is
Kaluza-Klein, the associated Harrison transformations are no longer
charging, one of them ($X_{11}$) being a five-dimensional gauge
transformation. On the other hand, the Ehlers transformation now
generates a five-dimensional rotation \cite{gisa} (note that
generation of rotation by sigma-model dualities is not possible at
all in four-dimensional theories).

To construct a new solution one has first to identify the target
space coordinates in terms of the seed solution, then to apply  the
above transformations to find new target space variables, and
finally to perform inverse dualization to get the metric and the
vector potential. However, to apply the most interesting
Harrison-Ehlers transformations we need to know  an explicit
transformation from the initial to the tilded potentials. To find it
in a concise form is still an open problem.

An alternative way consists in constructing a matrix representation
of the target space. First we have to reveal the nature of the
isotropy group $\HH\in G=G_{2(2)}$. This can be done as follows. One
looks for a representative of the coset $N(\Phi)\in G/\HH$ which
transforms by the global left multiplication by $g\in G$ and the
local right multiplication by $h(\Phi)\in\HH$:
\begin{equation}\label{N_trans}
N \to g N h(\Phi).
\end{equation}
Infinitesimally this reads
\begin{equation}\label{local}
X_M N(\Phi)=x_M N(\Phi) + q_M^\alpha(\Phi) N(\Phi)y_\alpha,
\end{equation}
where $X_M,\, M=1,\ldots,14$ are Killing vectors acting as
differential operators, $x_M$ are the corresponding matrices of
the $g_{2(2)}$ algebra,  $y_\alpha,\, \alpha=1,\ldots,6$ are
generators of the isotropy subalgebra $\HH$ and $q_M^\alpha(\Phi)$
are ``compensating'' functions. Consider the $7\times 7$ matrix
representation of $g_{2(2)}$ as Z-matrix (6.2) in \cite{gugu} with
omitted ``{\em i}'' to get the non-compact real form of the $g_2$
algebra. It is easy to find matrices $x_M$ corresponding to the
differential operators $X_M$. Solving the Eqs.(\ref{local})  for
the Borel subalgebra $M=1,\ldots,8$ with the assumption that for
these values of $M$ the functions $q_M^\alpha(\Phi)=0$, one
obtains the matrix $N(\Phi)$. Then applying (\ref{local}) for
$M=9,\ldots,14$ we find $q_M^\alpha(\Phi)$ together with the
corresponding $y_\alpha$. This calculation, requiring the explicit
form of the ``complicated'' generators $X_M$, was performed using
MAPLE. The resulting isotropy subalgebra $\HH$ turns out to be
spanned by
\begin{eqnarray}\label{y}
&&Y_1=X_8-X_9,\;Y_2=X_{11}-X_4,\;Y_3=3X_5+X_{12},\;
\nonumber\\
&&Y_4=3X_3+X_{10},\;Y_5=X_{2}+\frac1{\sqrt3}X_{13},\;Y_6=3X_1-X_{14},
\end{eqnarray}
which form $sl(2,R)\times sl(2,R)$. \par Using the matrix $N(\Phi)$
one can then pass to the gauge-independent matrix
$$M=N\eta N^{-1},$$ where $\eta$ is some
constant matrix  invariant under $\HH$ transformations $$h\eta
h^{-1}=\eta,\quad h\in \HH,$$ such that the target space metric reads
\begin{equation}\label{M}
    dl^2=\frac18 Tr\left(dM M^{-1}dM M^{-1}\right).
\end{equation}
The matrix $M$ by construction will be invariant under the local
action of $\HH$. Before giving its explicit form we would like to
go over to a more concise notation using the 2-dimensional covariance
of our construction with respect to the reduced dimensions $t,z$.

\section{2d-covariant representation}
The covariant parametrisation for dimensional reduction of
five-dimensional minimal supergravity with two commuting isometries
is \cite{ms,mizo1}
\begin{eqnarray}
ds_5^2 &=& \lambda_{ab}(dz^a + a_i^a dx^i)(dz^b + a_j^b dx^j) +
\tau^{-1} h_{ij} dx^i dx^j, \\
\hat A &=& \sqrt3 \psi_a dz^a + A_i dx^i
\end{eqnarray}
($a,b =0,1,\, z^0=t, z^1=z\,; i, j = 1, 2, 3$), where $\tau \equiv -
\det \lambda$. This is related to our previous parametrisation
arising from the two-step reduction (\ref{ds5}), (\ref{ds4}) by
\begin{equation}
\lambda =
 \left( \begin{array}{cc}
 - \mathrm{e}^{\phi+\xi}   + \mathrm{e}^{-2\phi} v_1^2 &
   \mathrm{e}^{-2\phi}v_1 \\
 \mathrm{e}^{-2\phi} v_1 &  \mathrm{e}^{-2\phi}
 \end{array} \right),
 \quad  a^0_i = - \omega_i,\quad a^1_i = C_i + v_1 \omega_i,\quad
\psi =
 \left( \begin{array}{cc}
  v_2/\sqrt3 + \kappa v_1\\ \kappa
 \end{array} \right).
\end{equation}
Dualization of $F_{ij} \equiv \partial_i A_j - \partial_j A_i$ gives
the scalar $\mu$ via
\begin{equation}\lb{dualmu}
\frac1{\sqrt3} F^{ij} = a^{aj} \partial^i \psi_a - a^{ai}
\partial^j \psi_a + \frac1{\tau \sqrt{h}} \epsilon^{ijk}
\left(
\partial_k \mu +   \epsilon^{ab} \psi_a \partial_k
\psi_b \right).
\end{equation}
This is related to previous quantities by
 $\mu=(u_2-\kappa
v_2)/\sqrt3$. Dualization of  $G^b_{ij} \equiv \partial_i a^b_j -
\partial_j a^b_i$ gives the two-vector $\omega^a$:
\begin{equation}\lb{dualom} \tau \lambda_{ab} G^{bij} =
\frac1{\sqrt{h}} \epsilon^{ijk} \left[
\partial_k \omega_a  - \psi_a \left( 3  \partial_k \mu +
  \epsilon^{bc} \psi_b
\partial_k \psi_c \right) \right],
\end{equation}
which is expressed as
\begin{equation}
\omega =
 \left( \begin{array}{cc}
 -\chi- \kappa v_2\left(\kappa v_1+2
 v_2/\sqrt{3}\right)/\sqrt3
 \\ u_1-\kappa^2 v_2/\sqrt3
 \end{array} \right).
\end{equation}
In this notation the target space metric (\ref{metric}) reads:
\begin{equation}\label{tarmet}
 dl^2 = \frac14 \mathrm{Tr} \left( \lambda^{-1} d\lambda \lambda^{-1}
d\lambda \right) + \frac14 \tau^{-2} d\tau^2 + \frac32 d\psi^T
\lambda^{-1} d\psi - \frac12\tau^{-1} V^T \lambda^{-1} V -
\frac32\tau^{-1} \left( d \mu +
  \epsilon^{ab} \psi_a d\psi_b \right)^2,
\end{equation}
(with $\epsilon^{01} = 1$), where the vector-valued one-form $V_a$
is
\begin{equation}
V_a = d\omega_a - \psi_a \left(  3   d\mu +  \epsilon^{bc} \psi_b
d\psi_c \right).
\end{equation}
The three coordinates $\mu$ and $\omega_a$ are cyclic.

The first eight (``simple'') Killing vectors $X_1,\ldots,X_8$
together with $X_{11}$ can be regrouped covariantly as follows. The
mixed tensor \be {M_a}^b =
2\lambda_{ac}\frac{\partial}{\partial\lambda_{cb}} +
\omega_a\frac{\partial}{\partial\omega_{b}} +
\delta_a^b\omega_c\frac{\partial}{\partial\omega_{c}} +
\psi_a\frac{\partial}{\partial\psi_{b}} +
\delta_a^b\mu\frac{\partial}{\partial\mu} \ee generates linear
transformations in the $(t,z)$ plane obeying the gl(2,R) subalgebra,
\be\lb{MM} \left[{M_a}^b,{M_c}^d\right] = \delta_c^b{M_a}^d -
\delta_a^d{M_c}^b. \ee Three mutually commuting operators are
associated with the three cyclic ``magnetic'' coordinates: \be
N^a =\frac{\partial}{\partial\omega_a}\,,\qquad  Q =
\frac{\partial}{\partial\mu}\,. \ee  Their commutation relations are
\bea
\left[{M_a}^b,N^c\right] &=& -(\delta_a^cN^b + \delta_a^bN^c)\,, \lb{MN}\\
\left[{M_a}^b,Q\right] &=& -\delta_a^bQ\,. \lb{MQ}\\
\left[N^a,N^b\right] &=& 0\,,   \lb{NN}\\
\left[Q, N^a\right] &=& 0\,. \lb{QN} \eea A doublet operator
associated with gauge transformations of the $\psi_a$ reads: \be R^a
= \left[\frac{\partial}{\partial\psi_a} +
3\mu\frac{\partial}{\partial\omega_a} -
 \epsilon^{ab}\psi_b\left(\frac{\partial}{\partial\mu} +
 \psi_c\frac{\partial}{\partial\omega_c}\right)\right]\,. \ee The corresponding
commutation relations are \bea
\left[{M_a}^b,R^c\right] &=& -\delta_a^cR^b \,, \lb{MR}\\
\left[N^a,R^b\right] &=& 0\,, \lb{NR}\\
\left[Q,R^a\right] &=& 3N^a\,, \lb{QR}\\
\left[R^a,R^b\right] &=& 2\epsilon^{ab}Q\,. \lb{RR} \eea The
correspondence with the previous operators is as follows: \bea
&&X_1=-N^0,\;\; X_2=N^1, \;\; X_3=\frac1{\sqrt3}  Q, \;\; X_4=
{M_1}^0, \;\;X_5=\frac1{\sqrt3}R^0,\;\;\nonumber\\
&&X_6={M_0}^0-{M_1}^1,\;\; X_7= 2{M_0}^0-{M_1}^1,\;\; X_8=- R^1 ,
\;\;X_{11}=-{M_0}^1.\eea Two more vectors $L_a:\,
X_{14}=3L_0,\,X_{13}=-\sqrt3   L_1 $ complete the subalgebra  $
sl(3,R)\in g_{2(2)}$: \bea
\left[{M_a}^b,L_c\right] &=& (\delta_c^bL_a + \delta_a^bL_c)\,,\lb{ML}\\
\left[N^a,L_b\right] &=& {M_b}^a\,, \lb{NL}\\
\left[L_a,L_b\right] &=& 0\,. \lb{LL} \eea Their commutation with
$Q$ gives another doublet: \be\lb{QL} \left[Q,L_a\right] = P_a
\,,\quad X_{12}=\sqrt3P_0,\;\; X_{9}=-  P_1.\ee Finally the algebra
$g_{2(2)}$ is closed with the singlet operator $T$, such that
\be\lb{RL} \left[R^a,L_b\right] = \delta^a_bT \,, \ee which is
identified as $X_{10}=-\sqrt3T$. The remaining commutation relations
are \begin{eqnarray}
\left[{M_a}^b,P_c\right] &=& \delta^b_cP_a \,, \nn\\
\left[{M_a}^b,T\right] &=& \delta^a_bT \,, \nn\\
\left[{N^a},P_b\right] &=& \delta^a_bQ \,, \nn\\
\left[{N^a},T\right] &=& R^a \,, \nn\\
\left[Q,P_a\right] &=& -2\epsilon_{ab}R^b\,, \nn\\
\left[Q,T\right] &=& {M_c}^c\,, \\
\left[{R^a},P_b\right] &=& -3{M^a}_b +\delta^a_b{M_c}^c \,,\nn\\
\left[{R^a},T\right] &=& 2\epsilon^{ab}P_b \,, \nn\\
\left[{L_a},P_b\right] &=& 0\,, \nn\\
\left[{L_a},T\right] &=& 0\,, \nn\\
\left[{P_a},P_b\right] &=& 2\epsilon_{ab}T\,, \nn\\
\left[{P_a},T\right] &=& 3L_a \,. \nn \end{eqnarray}

The most convenient way to find a matrix representative of the coset
consists in exponentiating the Borel subgroup. In our case this
amounts to using the one-parametric subgroups corresponding to
generators $X_1,\ldots, X_8$, or, in the two-covariant notation,
$N^a,\,Q,\,R^a$ and three independent components of ${M_a}^b$. The
latter, together with the $N^a$, generate representatives of the
vacuum coset $SL(3,R)/SL(2,R)$ \cite{ms} which is a submanifold of
the full target space. These representatives are of the form \be M_1
= {\cal V}^T_{\omega}M_0{\cal V}_{\omega}, \quad {\cal V}_\omega
=e^{\omega_a n^a},\ee where $M_0$ is the $7\times7$ matrix \be M_0 =
\left(\begin{array}{ccccc}
\lambda & 0 & 0 & 0 & 0 \\
0 & -\tau^{-1} & 0 & 0 & 0 \\
0 & 0 & \lambda^{-1} & 0 & 0 \\
0 & 0 & 0 & -\tau & 0 \\
0 & 0 & 0 & 0 & 1
\end{array}\right),
\ee and $\lambda$ and $\lambda^{-1}$ are $2\times 2$ blocks. Then
the full $G_{2(2)}/(SL(2,R)\times SL(2,R))$ coset matrix can be
constructed as \be\lb{fullM} M =  {\cal V}^T M_0{\cal V}, \quad
{\cal V}={\cal V}_\psi{\cal V}_\mu{\cal V}_\omega, \ee where \be
{\cal V}_\mu=e^{\mu q},\quad {\cal V}_\psi= \e^{\psi_a r^a}, \ee
with $\omega_a,\,\mu,\,\psi_a$ the target space coordinates and
$n^a,\,q,\,r^a$ the $7\times 7$ matrices of the $g_{2(2)}$ algebra
given in Appendix A. The computation gives the coset matrix in the
symmetrical block form \be M = \left(\begin{array}{ccc}
A & B & \sqrt2U \\
B^T & C & \sqrt2V \\
\sqrt2U^T & \sqrt2V^T & S
\end{array}\right),
\ee with \be
\begin{array}{l}
A = \left(\begin{array}{cc} \begin{array}{c}
\left[(1-y)\lambda + (2+x)\psi\psi^T  - \tau^{-1}\tom\tom^T\right.\\
\left.+\mu(\psi\psi^T\lambda^{-1}J - J\lambda^{-1}\psi\psi^T)\right]
\end{array} & \tau^{-1}\tom \\
\tau^{-1}\tom^T & -\tau^{-1}
\end{array}\right), \\
B = \left(\begin{array}{cc} (\psi\psi^T-\mu J)\lambda^{-1} -
\tau^{-1}\tom\psi^TJ &
\begin{array}{c}
\left[(-(1+y)\lambda J - (2+x)\mu +
\psi^T\lambda^{-1}\tom)\psi\right.
\\ \left. + (z - \mu J\lambda^{-1})\tom \right]
\end{array} \\
 \tau^{-1}\psi^TJ & -z
\end{array}\right), \\
C = \left(\begin{array}{cc} (1+x)\lambda^{-1} -
\lambda^{-1}\psi\psi^T\lambda^{-1} &
\lambda^{-1}\tom-J(z-\mu J\lambda^{-1})\psi\\
\tom^T\lambda^{-1} + \psi^T(z+\mu\lambda^{-1}J)J  &
\begin{array}{c}
\left[\tom^T\lambda^{-1}\tom - 2\mu\psi^T\lambda^{-1}\tom\right. \\
\left. -\tau(1+x-2y-xy+z^2) \right]
\end{array}
\end{array}\right), \\
U = \left(\begin{array}{c} (1+x-\mu J\lambda^{-1})\psi -
\mu\tau^{-1}\tom \\ \mu\tau^{-1}
\end{array}\right), \\
V = \left(\begin{array}{c}
(\lambda^{-1} + \mu\tau^{-1}J)\psi \\
\psi^T\lambda^{-1}\tom - \mu(1+x-z)
\end{array}\right), \\
S = 1+2(x-y)\,,
\end{array}
\ee where \be \tom = \omega - \mu\psi\,,\quad x =
\psi^T\lambda^{-1}\psi\,, \quad y = \tau^{-1}\mu^2\,, \quad z = y -
\tau^{-1}\psi^TJ\tom\,, \ee and $J$ is the $2\times 2$ matrix $$ J =
\left(\begin{array}{cc}
0 & 1 \\
-1 & 0
\end{array}\right).
$$

The matrix $M$ solves the equation \be\lb{J1} M^{-1}dM = {\cal J} =
16\eta^{MN}\bar{J}_Mj_N, \ee where \be\bar{J_M} = G_{AB}J_M^Ad\Phi^B
\ee are the one-forms dual to the Killing vectors $J_M =
J_M^A\partial/\partial\Phi^A$. In the present case, \bea {\cal J}
&=& \left({\bar{M}_b}^a - \frac13\delta_b^aTr\bar{M}\right){m_a}^b
+ \bar{N}^an^{aT} + \bar{L}_a{\ell_a}^T \nonumber \\
&& + \frac13\bigg(\bar{R}^ar^{aT} + \bar{P}_a{p_a}^T + \bar{Q}q^T +
\bar{T}t^T \bigg)\,. \eea

The action of $G_{2(2)}$  on the coset matrix is
\begin{equation}\label{transmat}
M' (\Phi^A)= {\cal C}^T M(\Phi^A) \;{\cal C},\quad  {\cal C}\in G,
\end{equation}
where $ {\cal C}$ is the exponential of some Lie algebra generator.
The strategy to generate a new solution consists in the
following steps. First, one must identify the target space
coordinates corresponding to the seed solution and form the matrix
$M(\Phi^A)$ as function of these variables (in the 2-covariant form
$\Phi^A= (\lambda_{ab}, \omega_a, \psi_a, \mu$). This involves solving the
duality equations (\ref{dualmu}), (\ref{dualom}) for the magnetic-type
potentials. Then one chooses a
transformation ${\cal C}$ of the desired type and computes the
transformed matrix $M' (\Phi^A)$ in terms of the same target space
variables. The new target space variables $\Phi'^A= (\lambda'_{ab},
\omega'_a, \psi'_a, \mu'$) can then be found by solving the set of
equations
\begin{equation}\label{primedvar}
M(\Phi'^A) = M' (\Phi^A).
\end{equation}
Finally, the metric functions and the vector potential of the new
solution must be calculated. This step also involves solving differential
equations to dualize back the potentials.

We have tested our formalism by generating the charged rotating
black hole solution of five-dimensional supergravity from the
Myers-Perry black hole. This is achieved by applying a combination
of an electric Harrison ($X_{12}$) transformation accompanied by the
corresponding gauge transformation in order to preserve asymptotic
flatness of the solution. The resulting solution can be transformed
to the previously known one by some coordinate transformation. This
is given in the Appendix B.

\section{Charged black ring with two parameters of rotation}
As an example of application of our formalism we derive here the
electrically charged version of the rotating black ring with two
parameters of rotation constructed by  Pomeransky and Senkov using
the inverse scattering technique \cite{posen}
\begin{eqnarray}
&&ds^2 = - \frac{H(y,x)}{H(x,y)} (dt + \Omega)^2 -
\frac{F(x,y)}{H(y,x)} d\phi^2 - 2\frac{J(x,y)}{H(y,x)} d\phi d\psi
\nonumber\\&&+ \frac{F(y,x)}{H(y,x)} d\psi^2 + \frac{2 k^2
H(x,y)}{(x-y)^2 (1-\nu)^2} \left[ \frac{dx^2}{G(x)} -
\frac{dy^2}{G(y)} \right],
\end{eqnarray}
where the variables ($t, x, y, \phi, \psi$) vary in the range
$-\infty < t < +\infty, -1 \leq x \leq 1, -\infty < y <-1, 0 \leq
(\phi, \psi) < 2 \pi,\;k,\,\nu,\,\lambda$ are parameters, and the
rotation one-form is two-component $\Omega = \Omega_\psi d\psi +
\Omega_\phi d\phi$. Explicitly,
\begin{equation} \label{omega}
\Omega  = - \frac{2 k \lambda \sqrt{(1 + \nu )^2 -
\lambda^2}}{H(y,x)} \left[ (1 - x^2) y \sqrt{\nu} d\psi + \frac{1 +
y}{1 - \lambda + \nu} [ 1 + \lambda - \nu + x^2 y \nu (1 - \lambda -
\nu) + 2 \nu x (1 - y) ] \, d\phi \right],
\end{equation}
and the functions $G, H, J, F$ are  (we use here the original
notation of \cite{posen}, not to be confused with our fields
$\lambda,x,y$):
\begin{eqnarray} \label{functions}
G(x) &=& (1 - x^2) (1 + \lambda x + \nu x^2),
\nonumber\\
H(x,y) &=& 1 + \lambda^2 - \nu^2 + 2 \lambda \nu (1 - x^2) y + 2 x
\lambda (1 - y^2 \nu^2) + x^2 y^2 \nu (1 - \lambda^2 - \nu^2),
\nonumber\\
J(x,y) &=& \frac{2 k^2 (1 - x^2) (1 - y^2) \lambda \sqrt{\nu}}{(x -
y) (1 - \nu)^2} \left[ 1 + \lambda^2 - \nu^2 + 2 (x + y) \lambda \nu
- x y \nu (1 - \lambda^2 - \nu^2) \right],
\\
F(x,y) &=& \frac{2 k^2}{(x - y)^2 (1 - \nu)^2} \Bigl\{ G(x) ( 1 -
y^2) \Bigl( [ (1 - \nu)^2 - \lambda^2 ] (1 + \nu) + y \lambda (1 -
\lambda^2 + 2 \nu - 3 \nu^2) \Bigr) + G(y) \Bigl[ 2 \lambda^2
\nonumber\\
&+& x \lambda [ (1 - \nu)^2 + \lambda^2 ] + x^2 [ (1 - \nu )^2 -
\lambda^2 ] (1 + \nu) + x^3 \lambda (1 - \lambda^2 - 3 \nu^2 + 2
\nu^3) - x^4 (1 - \nu) \nu (-1 + \lambda^2 + \nu^2) \Bigr] \Bigr\}.
\nonumber
\end{eqnarray}
Regularity of the black ring implies the inequalities
$0\leq\nu<1,\;2\sqrt{\nu}\leq\lambda<1+\nu.$ The mass and
angular momenta can be read out from the asymptotic expansion of the
metric:
\begin{equation}
M = \frac{3 k^2 \pi \lambda}{G_N (1 - \lambda + \nu )}, \quad J_\psi
= \frac{4 k^3 \pi \lambda \sqrt{\nu} \sqrt{(1 + \nu)^2 -
\lambda^2}}{G_N (1 - \nu)^2 (1 - \lambda + \nu)}, \quad J_\phi =
\frac{2 k^3 \pi \lambda (1 + \lambda - 6 \nu + \lambda \nu + \nu^2)
\sqrt{(1 + \nu)^2 - \lambda^2}}{G_N (1 - \nu)^2 (1 - \lambda +
\nu)^2},
\end{equation}
where $G_N$ is the Newton constant. This solution is free of conical
and Dirac string singularities.

We would like to endow this solution with an electric charge. This
may be done by applying our formalism with the reduction along
$(t,\psi),\;(t,\phi),$ or $t$ and a linear combination of $\psi$ and
$\phi$. We will here consider both choices $(t,\psi)$ and $(t,\phi)$,
and show that they lead to the same result. The target space potentials
corresponding to dimensional reduction with respect to $(t,\psi)$ are
\begin{eqnarray}\lb{seedpot}
\lambda_{00} &=& - \frac{H(y,x)}{H(x,y)}, \quad \lambda_{01} = -
\frac{H(y,x)}{H(x,y)} \Omega_\psi, \quad \lambda_{11} =
\frac{F(y,x)}{H(y,x)} - \frac{H(y,x)}{H(x,y)} \Omega_\psi^2, \quad
\tau = \frac{F(y,x)}{H(x,y)},
\\
a^0_\phi &=& \Omega_\phi - a^1_\phi \Omega_\psi , \qquad a^1_\phi =
- \frac{J(x,y)}{F(y,x)},
\end{eqnarray}
and the reduced three-dimensional metric is
\be
h_{ij}dx^idx^j =  \frac{2 k^2}{(1-\nu)^2(x-y)^2} \left[F(y,x)
\left(\frac{dx^2}{G(x)}-\frac{dy^2}{G(y)}\right)- \frac{2k^2G(x)G(y)}
{(x-y)^2}d\phi^2 \right].
\ee
The ``hatted'' potentials $\hat{\lambda}_{ab},\,\hat{\tau},\,
\hat{a}^a_{\psi}$ and three-dimensional metric $\hat{h}_{ij}$
corresponding to dimensional reduction with respect
to $(t,\phi)$ are obtained from (\ref{seedpot}) by making the exchange
$F(x,y) \leftrightarrow -F(y,x)$ and $\Omega_{\phi} \leftrightarrow
\Omega_{\psi}$.

To generate the electric potential $\bar{v}_2=\sqrt3\psi_0 $ one
must perform the Harrison transformation generated by
$P_0=X_{12}/\sqrt3$. It turns out, however, that this transformation
alone does not preserve asymptotic flatness of the solution. To fix
this, one must add the corresponding gauge transformation
$R^0=\sqrt3X_5$, so that the resulting charging transformation
${\cal C}=P_0+R^0 = (X_{12}+3X_5)/\sqrt3$ belongs to the
$SL(2,R)\times SL(2,R)$ isotropy subgroup (see Eq. (\ref{y})). The
corresponding one-parameter subgroup is given by the exponential
\begin{equation}
{\cal C}=\e^{\alpha (p^0+r_0)}= \left(\begin{array}{ccccccc}
c^2 & 0 & 0 & s^2 & 0 & 0 & \sqrt2sc \\
0 & c & 0 & 0 & 0 & s & 0 \\
0 & 0 & c & 0 & -s & 0 & 0 \\
s^2 & 0 & 0 & c^2 & 0 & 0 & \sqrt2sc \\
0 & 0 & -s & 0 & c & 0 & 0 \\
0 & s & 0 & 0 & 0 & c & 0 \\
\sqrt2sc & 0 & 0 & \sqrt2sc & 0 & 0 & c^2 + s^2
\end{array}\right) = {\cal C}^T,
\end{equation}
where $\alpha$ is a parameter, and $c \equiv \cosh\alpha$, $s \equiv
\sinh\alpha$. Applying this transformation to the seed matrix $M$ built
from the potentials (\ref{seedpot}) leads to the transformed matrix
\begin{equation}\label{charmat}
M' = {\cal C}^T M {\cal C},
\end{equation}
and extracting the transformed potentials we obtain
\bea\lb{transpot}
\tau' & = & D^{-1}\tau \,, \nonumber\\
\lambda'_{00} &=& D^{-2}\lambda_{00}\,, \nonumber\\
\lambda'_{01} & = & D^{-2}[c^3\lambda_{01} +
s^3\lambda_{00}\omega_0]\,,\nonumber \\
\omega'_0 &=& D^{-2}[c^3(c^2 + s^2 + 2s^2\lambda_{00})\omega_0 -
s^3(2c^2 + (c^2+s^2)\lambda_{00})\lambda_{01}]\,, \nonumber\\
\omega'_1 &=& \omega_1 + D^{-2}s^3[-c^3\lambda_{01}^2 +
s(2c^2-\lambda_{00})\lambda_{01}\omega_0 - c^3\omega_0^2]\,, \\
\psi'_0 &=&  D^{-1}sc(1+\lambda_{00})\,,\nonumber \\
\psi'_1 &=&  D^{-1}sc(c\lambda_{01}-s\omega_0)\,, \nonumber\\
\mu' &=&  D^{-1}sc(c\omega_0-s\lambda_{01})\,,\nonumber \eea where
\be D = c^2+s^2\lambda_{00} = 1 + s^2(1+\lambda_{00}) \ee is common
to the two reductions.

The seed potentials $\omega_a$ or $\hat{\omega}_a$ are obtained by
dualizing the $a^a_{\phi}$ or $\hat{a}^a_{\psi}$ via Eq.
(\ref{dualom}). Inspection of the relations (\ref{transpot}) shows
that it is not necessary to compute $\omega_1$, while the
computation of $\omega_0$ yields simply \be \omega_0(x,y) = -
\Omega_{\phi}(y,x)\,, \qquad  \hat\omega_0(x,y) =
\Omega_{\psi}(y,x)\,.\ee To write down the charged solution, there
remains to dualize back the $\omega'_a$ and $\mu'$ to the
$a'^a_{\phi}$ and $A'_{\phi}$. It is easy to show (without explicit
dualization) from the above relations, that $a'^1_{\phi} =
a^1_{\phi}$, while the transformed gravimagnetic and magnetic fields
are given by \be G'^{0i\phi} = c^3G^{0i\phi} +
s^3\bigg[-\omega_1G^{1i\phi} +
  \frac{\lambda_{00}^2}{\tau\sqrt{h}}\epsilon^{ij}
\partial_j\Omega_{\psi}\bigg]\,, \lb{G'}
\ee \be F'^{i\phi} = a'^{0\phi}\partial^i\psi'_0+
a'^{1\phi}\partial^i\psi'_1 -
\frac{\sqrt3sc}{D\tau\sqrt{h}}\epsilon^{ij}(c\partial_j\omega_0 +
s\lambda_{00}^2\partial_j\Omega_{\psi})\,,    \lb{F'} \ee
leading to
\bea
a'^0_{\phi}(x,y) &=& c^3a^0_{\phi}(x,y) + s^3a^0_{\psi}(y,x)\,, \nn\\
A'_{\phi}(x,y) &=& -cH(y,x)\Omega_{\phi}(x,y) -
sH(x,y)\Omega_{\psi}(y,x)\,, \eea for dimensional reduction relative
to $(t,\psi)$, and \bea \hat{a}'^0_{\psi}(x,y) &=&
c^3\hat{a}^0_{\psi}(x,y) - s^3\hat{a}^0_{\phi}(y,x)
\,, \nn\\
\hat{A}'_{\psi}(x,y) &=& -cH(y,x)\Omega_{\psi}(x,y) +
sH(x,y)\Omega_{\phi}(y,x) \,,\eea for dimensional reduction relative
to $(t,\phi)$. Both dimensional reductions lead to the same final
charged black ring metric \bea ds'^2 & = &
-D^{-2}\frac{H(y,x)}{H(x,y)}(dt+\Omega')^2 +
D\bigg[-\frac{F(x,y)}{H(y,x)}d\phi^2
 -2\frac{J(x,y)}{H(y,x)}d\phi d\psi\nonumber \\
&& + \frac{F(y,x)}{H(y,x)}d\psi^2 +
\frac{2k^2H(x,y)}{(1-\nu)^2(x-y)^2}\bigg(\frac{dx^2}{G(x)} -
\frac{dy^2}{G(y)}\bigg)\bigg]\,, \eea \be \Omega' =
(c^3\Omega_{\psi}(x,y) - s^3\Omega_{\phi}(y,x))d\psi +
(c^3\Omega_{\phi}(x,y) + s^3\Omega_{\psi}(y,x))d\phi\,, \ee  \bea
&&A' = \frac{\sqrt3sc}{DH(x,y)}\bigg[2\lambda(1-\nu)(x-y)(1-\nu xy)\,dt +\nn\\
&&+\bigg(sH(x,y)\Omega_{\phi}(y,x)-cH(y,x)\Omega_{\psi}(x,y)\bigg)d\psi\nn\\
&&-\bigg(sH(x,y)\Omega_{\psi}(y,x)+cH(y,x)\Omega_{\phi}(x,y)\bigg)d\phi\bigg],
\eea with \be D = 1 +
s^2\frac{2\lambda(1-\nu)(x-y)(1-\nu xy)}{H(x,y)}\,. \ee

This is to be compared with the charged black ring given in
\cite{EEF}, Sect. 4. A difference is that the authors of \cite{EEF}
start with a seed having an extra parameter (dipole charge), which
can be fine tuned so that Dirac-Misner strings are absent. Such
string singularities arise if the orbits of $\partial_{\psi}$
\footnote{The $\psi$ and $\phi$ of \cite{EEF} should be exchanged to
conform to the notations of \cite{posen}.} do not close off at
$x=1$. In the present case it is clear that both
$\Omega'_{\psi}(1,y)$ and $A'_{\psi}(1,y)$ are proportional to
$\Omega_{\phi}(y,1)$, which does not vanish, so that string
singularities are unavoidable. Specifically, \be \Omega'_{\psi}(1,y)
= -s^3\frac{4k\lambda}{\sqrt{(1+\nu)^2-\lambda^2}}\,. \ee To
eliminate this singularity one can apply a further transformation
depending on an extra parameter which can be adjusted to get rid of
the Dirac string \cite{preparation}.

\section{Conclusion}
In this paper we have presented a new solution-generating technique
for $D=5$ minimal supergravity based on the hidden symmetry $G_2$.
This opens the possibility of finding new families of solutions
possessing two commuting Killing symmetries. In this case the
bosonic equations of motion reduce to those of a three-dimensional
gravity coupled sigma-model on a symmetric space. Here we have
elaborated in detail the case of one time-like and one space-like
Killing vectors, leading to the $G_{2(2)}/(SL(2,R)\times SL(2,R))$
target space. The case of two space-like symmetries can be dealt
with along the same lines, the target space being instead
$G_{2(2)}/SO(4)$ . In four-dimensional theories these two cases are
usually interpreted as corresponding to stationary axisymmetric and
plane-wave space-times. In five dimensions one has more freedom to
choose a pair of two commuting symmetries, so one can use this
approach in wider classes of space-times. Moreover, for a similar
reason the generating symmetries can be more efficient. For
instance, in the four-dimensional case one cannot generate the Kerr
metric from the Schwarzschild metric using only sigma-model
symmetries. To achieve this, one must proceed to higher level
transformations \cite{kerr} belonging to the Geroch group associated
with the infinite extension of the sigma-model symmetries for
configurations depending only on two variables. In five dimensions,
rotating black holes can be generated from static black holes by
using only the sigma-model symmetries \cite{gisa}.

An interesting application can be expected for black rings in $D=5$
minimal supergravity. The most general (pure black ring) solutions
obtained so far are three-parametric: the charged black ring with
one rotation parameter \cite{EEF} and the doubly rotating uncharged
black ring \cite{posen}. It was conjectured that the generic
solution of this kind should contain five independent parameters: a
mass, an electric charge, a dipole charge, and two angular momenta.
 To obtain such a solution free
of conical and Dirac  singularities (if it exists) one should
incorporate primarily seven free parameters, with two extra
parameters to be fixed by imposing the regularity conditions. One
can show that out of the fourteen $G_{2(2)}$ transformations six
preserve the asymptotic behaviour of the black ring. From these six
three are gauge, while the other three generate a charge, a dipole
moment and an angular momentum. So from the three-parameter
unbalanced neutral black ring as given in \cite{er1} it is possible
in principle to generate a six-parameter solution, which should lead
to a four-parameter non-singular black ring \cite{preparation}. Here
we have given only a four-parameter solution (doubly rotating
charged black ring) which is plagued by the Dirac string
singularity.

An even farther reaching perspective consists in further reducing
our sigma model to two dimensions and constructing a
Belinski-Zakharov type integrable model or deriving B\"{a}klund
transformations. Formally this can be done in the same way as was
recently used
\cite{har1,im,tn,ko,pome1,ak,mi,tmy,ef,posen,ekri,yaza1} in the
purely vacuum sector, but the rank of matrices involved is increased
from three to seven (though in the block form) which makes the
problem technically more difficult.

\begin{acknowledgments}
The authors would like to thank Paul Sorba for numerous illuminating
discussions and helpful advice in dealing with the $g_2$ algebra.
D.G. is grateful to the LAPTH for hospitality in July 2007 while the
paper was written. C-M.C. would like to thank the LAPTH for
hospitality in August 2006 when this work was initiated. The work of
C-M.C. was supported by the National Science Council of the R.O.C.
under the grant NSC 96-2112-M-008-006-MY3, and in part by the
National Center of Theoretical Sciences and Center for Mathematics
and Theoretic Physics (NCU). T.W. acknowledges the use of SHARCNET
computing facilities (www.sharcnet.ca).

\end{acknowledgments}

\begin{appendix}

\section{Matrix representation}
Starting with the  real form of $g_2$ in the split octonion basis as
given in \cite{gugu}, it is straightforward to find the desired
representation for the generators used in the main text. Their
generic block decomposition is
\begin{equation}\label{jgen}
j = \left( \begin{array}{ccc}
  S & \tilde{V} & \sqrt2U \\
  - \tilde{U} & - S^T & \sqrt2 V \\
  \sqrt2 V^T &  \sqrt2 U^T & 0
\end{array}\right),
\end{equation}
where $S$ is a $3\times3$ matrix, $U$ and $V$ are 3-component column
matrices, and $\tilde{U}$, $\tilde{V}$ are the $3\times3$ dual
matrices $\tilde{U}_{ij} = \epsilon_{ijk}U_k$. The matrices
${m_a}^b$, $n^a$ and $\ell_a$ generating SL(3,R) are of type $S$,
the corresponding $3\times3$ blocks being \bea S_{{m_0}^0}\!\!
&=&\!\! \left(\begin{array}{ccc}1&0&0\\0&0&0\\0&0&-1
\end{array}\right),\;\;
S_{{m_0}^1} =
\left(\begin{array}{ccc}0&1&0\\0&0&0\\0&0&0\end{array}\right) ,\;\;
S_{{m_1}^0} =
\left(\begin{array}{ccc}0&0&0\\1&0&0\\0&0&0\end{array}\right) ,\;\;
S_{{m_1}^1} =
\left(\begin{array}{ccc}0&0&0\\0&1&0\\0&0&-1\end{array}\right) ,\;\;
\nonumber\\
S_{n^0}\! \!&=&\!\! \left(\begin{array}{ccc}0&0&0\\0&0&0\\-1&0&0
\end{array}\right),\;\;
S_{n^1} =
\left(\begin{array}{ccc}0&0&0\\0&0&0\\0&-1&0\end{array}\right),\;\;
S_{\ell_0} =
\left(\begin{array}{ccc}0&0&1\\0&0&0\\0&0&0\end{array}\right),\;\;
S_{\ell_1} =
\left(\begin{array}{ccc}0&0&0\\0&0&1\\0&0&0\end{array}\right). \eea
The matrices $p_a$ and $q$ are of type $U$,  the corresponding
$1\times3$ blocks being \be U_{p_0} =  \left(\begin{array}{c} 1 \\ 0 \\
0
\end{array}\right),\;\;
U_{p_1} =  \left(\begin{array}{c} 0 \\ 1 \\ 0
\end{array}\right),\;\;
U_{q} =   \left(\begin{array}{c} 0 \\ 0 \\ -1
\end{array}\right).
\ee The matrices $r^a$ and $t$ are of type $V$,  the corresponding
$1\times3$ blocks being \be V_{r^0} =  \left(\begin{array}{c} 1 \\ 0 \\
0
\end{array}\right),\;\;
V_{r^1} =  \left(\begin{array}{c} 0 \\ 1 \\ 0
\end{array}\right),\;\;
V_{t} =   \left(\begin{array}{c} 0 \\ 0 \\ 1
\end{array}\right).
\ee Note that  the transposed matrices $j_A^T$ are related to the
original matrices $j_A$ by \be\lb{jtk} j_A^T = -Kj_AK\,, \ee where
the involution $K$ has the block structure \be K=
\left(\begin{array}{ccc}0&1&0\\1&0&0\\0&0&-1\end{array}\right)\,.
\ee

The real matrix representing the coset $G_{2(2)}/(SL(2,R)\times
SL(2,R))$ may be chosen to have the symmetrical block structure
\be\lb{Mblock} M = \left(\begin{array}{ccc}
A & B & \sqrt2U \\
B^T & C & \sqrt2V \\
\sqrt2U^T & \sqrt2V^T & S
\end{array}\right),
\ee where $A$ and $C$ are symmetrical $3\times3$ matrices, $B$ is a
$3\times3$ matrix, $U$ and $V$ are 3-component column matrices, and
$S$ is a scalar such that the inverse matrix reads \be M^{-1} =KMK=
\left(\begin{array}{ccc}
C & B^T & -\sqrt2V \\
B & A & -\sqrt2U \\
-\sqrt2V^T & -\sqrt2U^T & S
\end{array}\right).\ee

\section{Black holes}
Here we illustrate the application of our technique to generate the
charged rotating non-BPS black holes of five-dimensional supergravity,
starting with the five-dimensional vacuum Kerr metric \cite{mype}:
\begin{equation}
ds^2 = - dt^2 + \frac{\rho^2}{4\Delta} dx^2 + \rho^2 d\theta^2 + (x
+ a^2) \sin^2\theta d\phi^2 + (x + b^2) \cos^2\theta d\psi^2 +
\frac{r_0^2}{\rho^2} \left( dt + a \sin^2\theta d\phi + b
\cos^2\theta d\psi \right)^2,
\end{equation}
where
\begin{equation}
\rho^2 = x + a^2 \cos^2\theta + b^2 \sin^2\theta, \qquad \Delta = (x
+ a^2)(x + b^2) - r_0^2 x.
\end{equation}

Choosing  $z=\psi$ as the reduced direction,  we find the following
identification with our variables:
\begin{eqnarray}
\lambda_{00} &=& - 1 + \frac{r_0^2}{\rho^2}, \quad \lambda_{01} =
\frac{r_0^2}{\rho^2} b \cos^2\theta, \quad \lambda_{11} = (x + b^2)
\cos^2\theta + \frac{r_0^2}{\rho^2} b^2 \cos^4\theta, \quad \tau =
\left( x + b^2 - r_0^2 + \frac{r_0^2}{\rho^2} a^2 \cos^2\theta
\right) \cos^2\theta,
\nonumber\\
a^0_\phi &=& - \tau^{-1} \frac{r_0^2}{\rho^2} (x + b^2) a
\sin^2\theta \cos^2\theta, \qquad a^1_\phi = \tau^{-1}
\frac{r_0^2}{\rho^2} a b \sin^2\theta \cos^2\theta,
\end{eqnarray}
and the invariant three-metric
\begin{equation}
h_{ij} dx^i dx^j = \tau \left( \frac{\rho^2}{4 \Delta} dx^2 + \rho^2
d\theta^2 + \frac{\Delta}{\tau} \sin^2\theta \cos^2\theta d\phi^2
\right), \qquad \sqrt{h} = \frac12 \tau \rho^2 \sin\theta
\cos\theta.
\end{equation}
The dualization of the vector fields gives
\begin{equation}
\omega_0 = - \frac{r_0^2}{\rho^2} a \cos^2\theta, \qquad \omega_1= -
\frac{r_0^2}{\rho^2} a b \cos^4\theta.
\end{equation}

The action of our charge-generating transformation with parameter
$\alpha$ ($c=\cosh \alpha,\, s=\sinh\alpha$) on this neutral seed
leads to the transformed potentials according to (\ref{transpot}).
Performing the inverse dualization, we obtain the  charged black
hole solution
\begin{eqnarray}
ds'^2 &=& - D^{-2} \left(1 - \frac{r_0^2}{\rho^2} \right) (dt +
\Omega')^2 + D \left[ \frac{\rho^2 dx^2}{4 \Delta} + \rho^2
d\theta^2 + \left( x + a^2 + \frac{r_0^2 a^2}{\rho^2 - r_0^2}
\sin^2\theta \right) \sin^2 \theta d\phi^2 \right.
\nonumber\\
& & \left. + 2\frac{r_0^2 a b}{\rho^2 - r_0^2} \sin^2\theta
\cos^2\theta d\phi d\psi + \left( x + b^2 + \frac{r_0^2 b^2}{\rho^2
- r_0^2} \cos^2\theta \right) \cos^2\theta d\psi^2 \right],
\\
\Omega' &=& -r_0^2 \left[ \left( \frac{c^3 a}{\rho^2 - r_0^2} +
\frac{s^3 b}{\rho^2} \right) \sin^2\theta d\phi + \left( \frac{c^3
b}{\rho^2 - r_0^2} + \frac{s^3 a}{\rho^2} \right) \cos^2\theta d\psi
\right],
\\
A' &=& \sqrt3 s c D^{-1} \frac{r_0^2}{\rho^2} \left[ dt + (c a + s
b) \sin^2\theta d\phi + (c b + s a) \cos^2\theta d\psi \right],
\end{eqnarray}
with
\begin{equation}
D = 1 + s^2 \frac{r_0^2}{\rho^2}.
\end{equation}
The same solution is obtained if reduction is carried out with
respect to the angular variable $\phi$ instead of $\psi$. Note that
it is regular on the polar axis  $\sin\theta = 0$.

This solution should be compared with the solution for the charged
rotating black hole first given in \cite{CY} and rederived, in a different
parametrization, in \cite{cve}. The comparison with \cite{CY} is
straightforward. The solution presented in \cite{cve} is, for $g = 0$,
\begin{eqnarray}
d \bar s^2 &=& - dt^2 - \frac{2 \bar q}{\bar \rho^2} \bar \nu (dt -
\bar \omega) + \frac{\bar f}{\bar \rho^4} (dt - \bar \omega)^2 +
\frac{\bar \rho^2 r^2}{\bar \Delta} dr^2 + \bar \rho^2 d\theta^2 +
(r^2 + \bar a^2) \sin^2\theta d\phi^2 + (r^2 + \bar b^2)
\cos^2\theta d\psi^2,
\\
\bar A &=& \frac{\sqrt3 \bar q}{\bar \rho^2} (dt - \bar \omega),
\end{eqnarray}
where
\begin{eqnarray}
\bar \nu &=& \bar b \sin^2\theta d\phi + \bar a \cos^2\theta d\psi,
\qquad \bar \omega = \bar a \sin^2\theta d\phi + \bar b \cos^2\theta
d\psi, \qquad \bar f = 2 \bar m \bar \rho^2 - \bar q^2,
\\
\bar \Delta &=& (r^2 + \bar a^2) (r^2 + \bar b^2) + \bar q^2 + 2
\bar a \bar b \bar q - 2 \bar m r^2, \qquad \bar \rho^2 = r^2 + \bar
a^2 \cos^2\theta + \bar b^2 \sin^2\theta.
\end{eqnarray}
The metrics $ds'^2$ and $d\bar{s}^2$ are related by the following
coordinate and parameter transformation:
\begin{equation}
r^2 = x + s^2 (r_0^2 - a^2 - b^2) - 2 a b s c, \qquad 2 \bar m = (1
+ 2 s^2) r_0^2, \qquad \bar q = - s c r_0^2, \qquad \bar a = - c a -
s b, \qquad \bar b = - c b - s a,
\end{equation}
implying
\begin{equation}
\bar \rho^2 = D \rho^2 = \rho^2 + s^2 r_0^2, \qquad \bar \Delta =
\Delta.
\end{equation}
Comparing then the electromagnetic potentials, we find $\bar A = - A'$, so
the two solutions are identical under charge conjugation (or a simultaneous
sign change of $t$, $\phi$ and $\psi$).
\end{appendix}



\begin{references}

\bibitem{er}
  R.~Emparan and H.~S.~Reall,
  ``A rotating black ring in five dimensions,''
  Phys.\ Rev.\ Lett.\  {\bf 88}, 101101 (2002)
  [arXiv:hep-th/0110260].

\bibitem{er1}
  R.~Emparan and H.~S.~Reall,
  ``Black rings,''
  Class.\ Quant.\ Grav.\  {\bf 23}, R169 (2006)
  [arXiv:hep-th/0608012].

\bibitem{cre}
  E.~Cremmer,
  ``Supergravities in 5 dimensions,''
  in {\it Superspace and Supergravity},
  Eds. S.~W.~Hawking and M.~Rocek
  (Cambridge Univ. Press, 1981) 331; 267.

\bibitem{chani}
  A.~H.~Chamseddine and H.~Nicolai,
  ``Coupling the SO(2) supergravity through dimensional reduction,''
  Phys.\ Lett.\  B {\bf 96}, 89 (1980).

\bibitem{gaun}
  J.~P.~Gauntlett, J.~B.~Gutowski, C.~M.~Hull, S.~Pakis and H.~S.~Reall,
  ``All supersymmetric solutions of minimal supergravity in five dimensions,''
  Class.\ Quant.\ Grav.\  {\bf 20}, 4587 (2003)
  [arXiv:hep-th/0209114].

\bibitem{elvasu}
  H.~Elvang, R.~Emparan, D.~Mateos and H.~S.~Reall,
  ``A supersymmetric black ring,''
  Phys.\ Rev.\ Lett.\  {\bf 93}, 211302 (2004)
  [arXiv:hep-th/0407065].

\bibitem{mizo1}
  S.~Mizoguchi and N.~Ohta,
  ``More on the similarity between D = 5 simple supergravity and M theory,''
  Phys.\ Lett.\  B {\bf 441}, 123 (1998)
  [arXiv:hep-th/9807111].

\bibitem{mizo2}
  S.~Mizoguchi and G.~Schr\"{o}der,
  ``On discrete U-duality in M-theory,''
  Class.\ Quant.\ Grav.\  {\bf 17}, 835 (2000)
  [arXiv:hep-th/9909150].

\bibitem{mype}
  R.~C.~Myers and M.~J.~Perry,
  ``Black holes in higher dimensional space-times,''
  Annals Phys.\  {\bf 172}, 304 (1986).

\bibitem{cve}
  M.~Cveti\u{c}, H.~L\"u, and C.~N.~Pope,
  ``Charged Kerr-de Sitter black holes in five dimensions,''
  Phys.\ Lett.\  B {\bf 598}, 273 (2004)
  [arXiv:hep-th/0406196];
\\
  M.~Cveti\u{c}, H.~L\"u, and C.~N.~Pope,
  ``Charged rotating black holes in five dimensional $U(1)^3$ gauged N = 2
  supergravity,''
  Phys.\ Rev.\  D {\bf 70}, 081502 (2004)
  [arXiv:hep-th/0407058];
\\
  Z.~W.~Chong, M.~Cveti\u{c}, H.~L\"u and C.~N.~Pope,
  ``General non-extremal rotating black holes in minimal five-dimensional
  gauged supergravity,''
  Phys.\ Rev.\ Lett.\  {\bf 95}, 161301 (2005)
  [arXiv:hep-th/0506029].

\bibitem{gmt}
  J.~P.~Gauntlett, R.~C.~Myers and P.~K.~Townsend,
  ``Black holes of D = 5 supergravity,''
  Class.\ Quant.\ Grav.\  {\bf 16}, 1 (1999)
  [arXiv:hep-th/9810204].

\bibitem{kunz}
  J.~Kunz and F.~Navarro-Lerida,
  ``Non-uniqueness, counterrotation, and negative horizon mass of
  Einstein-Maxwell-Chern-Simons black holes,''
  Mod.\ Phys.\ Lett.\  A {\bf 21}, 2621 (2006)
  [arXiv:hep-th/0610075].

\bibitem{cjlp}
  E.~Cremmer, B.~Julia, H.~L\"u and C.~N.~Pope,
  ``Higher-dimensional origin of D = 3 coset symmetries,''
  arXiv:hep-th/9909099.

\bibitem{cjlp1}
  E.~Cremmer, B.~Julia, H.~L\"u and C.~N.~Pope,
  ``Dualization of dualities. I,''
  Nucl.\ Phys.\  B {\bf 523}, 73 (1998)
  [arXiv:hep-th/9710119];
\\
  E.~Cremmer, B.~Julia, H.~L\"u and C.~N.~Pope,
  ``Dualization of dualities. II: Twisted self-duality of doubled fields and
  superdualities,''
  Nucl.\ Phys.\  B {\bf 535}, 242 (1998)
  [arXiv:hep-th/9806106].

\bibitem{pos1}
  M.~Possel,
  ``Hidden symmetries in five-dimensional supergravity,''
   PhD Thesis, Hamburg, 2003.

\bibitem{pos}
  M.~Possel and S.~Silva,
  ``Hidden symmetries in minimal five-dimensional supergravity,''
  Phys.\ Lett.\  B {\bf 580}, 273 (2004)
  [arXiv:hep-th/0310256].

\bibitem{Gilmore}
  R.~Gilmore,
  {\it Lie Groups, Lie Algebras and Some of Their Applications},
  John Wiley \& Sons, 1974.

\bibitem{mizo3}
  S.~Mizoguchi, K.~Mohri and Y.~Yamada,
  ``Five-dimensional supergravity and hyperbolic Kac-Moody algebra $G_2^H$,''
  Class.\ Quant.\ Grav.\  {\bf 23}, 3181 (2006)
  [arXiv:hep-th/0512092].

\bibitem{eh}
  J. Ehlers,
  in {\em Les Theories Relativistes de la Gravitation},
  CNRS, Paris, 1959, p. 275.

\bibitem{ms}
  D.~Maison,
  ``Ehlers-Harrison type transformations for Jordan's extended theory of
  gravitation,''
  Gen.\ Rel.\ Grav.\  {\bf 10}, 717 (1979).

\bibitem{ern}
  F.~J.~Ernst,
  ``New formulation of the axially symmetric gravitational field problem,''
  Phys.\ Rev.\  {\bf 167}, 1175 (1968);
\\
  F.~J.~Ernst,
  ``New formulation of the axially symmetric gravitational field problem. II,''
  Phys.\ Rev.\  {\bf 168}, 1415 (1968);
\\
W.~Kinnersley,
  ``Generation of stationary Einstein-Maxwell fields,''
  J.\ Math.\ Phys.\  {\bf 14}, 651 (1973);
\\
  W.~Kinnersley,
  ``Symmetries of the stationary Einstein-Maxwell field equations. I,''
  J.\ Math.\ Phys.\  {\bf 18}, 1529 (1977).

\bibitem{nk}
  G.~Neugebauer and D.~Kramer,
  ``A method for the construction of stationary Einstein-Maxwell fields. (in
  german),''
  Ann. der Physik (Leipzig) {\bf 24}, 62 (1969).

\bibitem{mg}
  P.~O.~Mazur,
  ``A relationship between the electrovacuum Ernst equations and nonlinear
  sigma model,''
  Acta Phys.\ Polon.\  B {\bf 14} (1983) 219.
\\
  A.~Eris, M.~G\"urses, and A.~Karasu,
  ``Symmetric space property and an inverse scattering formulation of the SAS
  Einstein-Maxwell field equations,''
  J.\ Math.\ Phys.\  {\bf 25} (1984) 1489.

\bibitem{gk}
  D.~V.~Gal'tsov and O.~V.~Kechkin,
  ``Ehlers-Harrison type transformations in dilaton - axion gravity,''
  Phys.\ Rev.\  D {\bf 50}, 7394 (1994)
  [arXiv:hep-th/9407155].

\bibitem{g}
  D.~V.~Gal'tsov,
  ``Integrable systems in stringy gravity,''
  Phys.\ Rev.\ Lett.\  {\bf 74}, 2863 (1995)
  [arXiv:hep-th/9410217].

\bibitem{diak}
  D.~V.~Gal'tsov and O.~V.~Kechkin,
  ``Matrix dilaton - axion for heterotic string in three-dimensions,''
  Phys.\ Lett.\  B {\bf 361}, 52 (1995)
  [arXiv:hep-th/9507164].

\bibitem{cg}
  G.~Cl\'ement and D.~V.~Gal'tsov,
  ``Stationary BPS solutions to dilaton-axion gravity,''
  Phys.\ Rev.\  D {\bf 54}, 6136 (1996)
  [arXiv:hep-th/9607043].

\bibitem{gl}
  D.~V.~Gal'tsov and P.~S.~Letelier,
  ``Ehlers-Harrison transformations and black holes in dilaton-axion gravity
  with multiple vector fields,''
  Phys.\ Rev.\  D {\bf 55}, 3580 (1997)
  [arXiv:gr-qc/9612007].

\bibitem{gsh}
  D.~V.~Gal'tsov and S.~A.~Sharakin,
  ``Matrix Ernst potentials for EMDA with multiple vector fields,''
  Phys.\ Lett.\  B {\bf 399}, 250 (1997)
  [arXiv:hep-th/9702039].

\bibitem{ch}
  C.~M.~Chen, D.~V.~Gal'tsov, K.~I.~Maeda and S.~A.~Sharakin,
  ``SL(4,R) generating symmetry in five-dimensional gravity coupled to dilaton
  and three-form,''
  Phys.\ Lett.\  B {\bf 453}, 7 (1999)
  [arXiv:hep-th/9901130].

\bibitem{ju}
  B. Julia,
  in {\it Unified field theories in more than 4 dimensions},
  V. De Sabbata, E. Schmutzer (eds.), Singapore, WS 1983.

\bibitem{bgm}
  P.~Breitenlohner, D.~Maison and G.~W.~Gibbons,
  ``Four-dimensional black holes from Kaluza-Klein theories,''
  Commun.\ Math.\ Phys.\  {\bf 120}, 295 (1988).

\bibitem{bm}
  P.~Breitenlohner and D.~Maison,
  ``On nonlinear sigma-models arising in (super-)gravity,''
  Commun.\ Math.\ Phys.\  {\bf 209}, 785 (2000)
  [arXiv:gr-qc/9806002].

\bibitem{sen2}
  A.~Sen,
  ``Strong - weak coupling duality in three-dimensional string theory,''
  Nucl.\ Phys.\  B {\bf 434}, 179 (1995)
  [arXiv:hep-th/9408083].

\bibitem{elva2}
  H.~Elvang,
  ``A charged rotating black ring,''
  Phys.\ Rev.\  D {\bf 68}, 124016 (2003)
  [arXiv:hep-th/0305247].

\bibitem{yaza}
  S.~S.~Yazadjiev,
  ``Completely integrable sector in 5D Einstein-Maxwell gravity and derivation
  of the dipole black ring solutions,''
  Phys.\ Rev.\  D {\bf 73}, 104007 (2006)
  [arXiv:hep-th/0602116];
\\
  S.~S.~Yazadjiev,
  ``Solution generating in 5D Einstein-Maxwell-dilaton gravity and derivation
  of dipole black ring solutions,''
  JHEP {\bf 0607}, 036 (2006)
  [arXiv:hep-th/0604140];
\\
  S.~S.~Yazadjiev,
  ``Rotating dyonic dipole black rings: exact solutions and thermodynamics,''
  Gen.\ Rel.\ Grav.\  {\bf 39}, 601 (2007)
  [arXiv:hep-th/0607101].

\bibitem{beruf}
  V.~Belinsky and R.~Ruffini,
  ``On axially symmetric soliton solutions of the coupled scalar vector tensor
  equations in general relativity,''
  Phys.\ Lett.\  B {\bf 89}, 195 (1980).

\bibitem{har1}
  T.~Harmark,
  ``Stationary and axisymmetric solutions of higher-dimensional general
  relativity,''
  Phys.\ Rev.\  D {\bf 70}, 124002 (2004)
  [arXiv:hep-th/0408141].

\bibitem{im}
  H.~Iguchi and T.~Mishima,
  ``Solitonic generation of five-dimensional black ring solution,''
  Phys.\ Rev.\  D {\bf 73}, 121501 (2006)
  [arXiv:hep-th/0604050];
\\
  S.~Tomizawa, H.~Iguchi and T.~Mishima,
  ``Relationship between solitonic solutions of five-dimensional Einstein
  equations,''
  Phys.\ Rev.\  D {\bf 74}, 104004 (2006)
  [arXiv:hep-th/0608169];
\\
  H.~Iguchi and T.~Mishima,
  ``Black di-ring and infinite nonuniqueness,''
  Phys.\ Rev.\  D {\bf 75}, 064018 (2007)
  [arXiv:hep-th/0701043].

\bibitem{tn}
  S.~Tomizawa and M.~Nozawa,
  ``Vaccum solutions of five-dimensional Einstein equations generated by
  inverse scattering method. II: Production of black ring solution,''
  Phys.\ Rev.\  D {\bf 73}, 124034 (2006)
  [arXiv:hep-th/0604067].

\bibitem{ko}
  T.~Koikawa,
  ``Infinite number of soliton solutions to 5-dimensional vacuum Einstein
  equation,''
  Prog.\ Theor.\ Phys.\  {\bf 114}, 793 (2005)
  [arXiv:hep-th/0501248].

\bibitem{pome1}
  A.~A.~Pomeransky,
  ``Complete integrability of higher-dimensional Einstein equations with
  additional symmetry, and rotating black holes,''
  Phys.\ Rev.\  D {\bf 73}, 044004 (2006)
  [arXiv:hep-th/0507250].

\bibitem{ak}
  T.~Azuma and T.~Koikawa,
  ``Infinite number of stationary soliton solutions to five-dimensional vacuum
  Einstein equation,''
  Prog.\ Theor.\ Phys.\  {\bf 116}, 319 (2006)
  [arXiv:hep-th/0512350].

\bibitem{mi}
  T.~Mishima and H.~Iguchi,
  ``New axisymmetric stationary solutions of five-dimensional vacuum  Einstein
  equations with asymptotic flatness,''
  Phys.\ Rev.\  D {\bf 73}, 044030 (2006)
  [arXiv:hep-th/0504018].

\bibitem{tmy}
  S.~Tomizawa, Y.~Morisawa and Y.~Yasui,
  ``Vacuum solutions of five dimensional Einstein equations generated by
  inverse scattering method,''
  Phys.\ Rev.\  D {\bf 73}, 064009 (2006)
  [arXiv:hep-th/0512252].

\bibitem{ef}
  H.~Elvang and P.~Figueras,
  ``Black Saturn,''
  arXiv:hep-th/0701035.

\bibitem{posen}
  A.~A.~Pomeransky and R.~A.~Sen'kov,
  ``Black ring with two angular momenta,''
  arXiv:hep-th/0612005.

\bibitem{ekri}
  J.~Evslin and C.~Krishnan,
  ``The black di-ring: an inverse scattering construction,''
  arXiv:0706.1231 [hep-th].

\bibitem{yaza1}
  S.~S.~Yazadjiev,
  ``Black Saturn with dipole ring,''
  arXiv:0705.1840 [hep-th].

\bibitem{dimred}
  S.~Ferrara, R.~R.~Khuri and R.~Minasian,
  ``M-Theory on a Calabi-Yau manifold,''
  Phys.\ Lett.\  B {\bf 375}, 81 (1996)
  [arXiv:hep-th/9602102];
\\
  S.~Ferrara, R.~Minasian and A.~Sagnotti,
  ``Low-energy analysis of $M$ and $F$ theories on Calabi-Yau
  threefolds,''
  Nucl.\ Phys.\  B {\bf 474}, 323 (1996)
  [arXiv:hep-th/9604097].

\bibitem{5to3}
G. Cl\'ement, in preparation.

\bibitem{crack}
T.~Wolf, "Applications of CRACK in the Classification of Integrable
Systems", CRM Proceedings and Lecture Notes, {\bf 37} (2004) 283 --
300, http://lie.math.brocku.ca/twolf/papers/crm.ps    (or crm.dvi)

\bibitem{gisa} S.~Giusto and A.~Saxena,
``Stationary axisymmetric solutions of five-dimensional gravity'',
[arXiv:0705.4484].


\bibitem{gugu}
  M.~Gunaydin and F.~Gursey,
  ``Quark structure and octonions,''
  J.\ Math.\ Phys.\  {\bf 14}, 1651 (1973).

\bibitem{EEF}
  H.~Elvang, R.~Emparan and P.~Figueras,
  ``Non-supersymmetric black rings as thermally excited supertubes,''
  JHEP {\bf 0502}, 031 (2005)
  [arXiv:hep-th/0412130].

\bibitem{preparation} A.~Bouchareb, G.~Cl\'ement, C-M.~Chen,
D.~V.~Gal'tsov and N.~G.~Scherbluk, work in progress.

\bb{kerr} G. Cl\'ement, ``From Schwarzschild to Kerr: Generating
spinning Einstein-Maxwell fields from static fields", Phys.\ Rev.\ D
{\bf 57}, 4885 (1998)
  [arXiv:gr-qc/9710109].

\bibitem{CY} M. Cveti\u{c} and D. Youm, "General rotating
five-dimensional black holes of toroidally compactified heterotic string",
Nucl. Phys. B {\bf 476}, 118 (1996) [arXiv:hep-th/9603100].

\end{references}
\end{document}